\documentclass[a4paper,fleqn,usenatbib]{mnras}

\pdfoutput=1

\usepackage[dvipsnames,usenames]{color}

\usepackage{colortbl}

\usepackage{mathptmx}

\usepackage{etoolbox}

\usepackage{graphicx}

\usepackage{amsmath}

\usepackage{natbib}

\newcommand{\etal}{{et al}\/.}

\pubyear{2016}

\newtoggle{highq}

\toggletrue{highq}

\newcommand{\magphys}{{\sc MagPhys}}

\newcommand{\hatlas}{{\it H}-ATLAS}

\newcommand{\cre}{CR{\it e}}

\begin{document}

\title[Low-frequency radio luminosity--SFR relation]{LOFAR/H-ATLAS: The low-frequency radio luminosity -- star-formation rate relation}

\author[G.\ G\"urkan \etal]
       {G.\ G\"urkan$^{1,2*}$, M.J.\ Hardcastle$^1$,
         D.J.B.\ Smith$^1$, P.N.\ Best$^{3}$, N. Bourne$^{3}$, G.
         Calistro-Rivera$^{4}$, \newauthor G. Heald$^{2}$,
         M.J.\ Jarvis$^{5,6}$, I. Prandoni$^{7}$, H. J. A.
         R\"ottgering$^{4}$, J. Sabater$^{3}$, T.
         Shimwell$^{4}$, \newauthor C. Tasse$^{8,9}$ and W.L.\ Williams$^1$\\ 
$^1$ Centre for Astrophyics Research, School of Physics, Astronomy and Mathematics, University of Hertfordshire, College Lane, Hatfield AL10 9AB, UK\\
$^2$ CSIRO Astronomy and Space Science, 26 Dick Perry Avenue, Kensington, Perth, 6151, WA, Australia\\  
$^3$ Institute for Astronomy, University of Edinburgh, Royal Observatory, Blackford Hill, Edinburgh, EH9 3HJ, UK\\
$^4$ Leiden Observatory, Leiden University, PO Box 9513, 2300 RA Leiden, the Netherlands\\
$^5$ Oxford Astrophysics, Denys Wilkinson Building, Keble Road, Oxford OX1 3RH, UK\\
$^6$ Physics Department, University of the Western Cape, Private Bag X17, Bellville 7535, South Africa\\
$^7$ INAF-Istituto di Radioastronomia, Via P. Gobetti 101, Bologna, I-40129\\
$^8$ GEPI, Observatoire de Paris, CNRS, Universite Paris Diderot, 5 place Jules Janssen, 92190 Meudon, France\\
$^9$ Department of Physics \& Electronics, Rhodes University, PO Box 94, Grahamstown, 6140, South Africa\\
$^*$ gulay.gurkan.g@gmail.com\\}

\maketitle

\begin{abstract}

Radio emission is a key indicator of star-formation activity in
 galaxies, but the radio luminosity-star formation relation has to date been studied almost exclusively at frequencies of 1.4 GHz or above. At lower radio frequencies the effects of thermal radio emission are greatly reduced, and so we would expect the radio emission observed to be completely dominated by synchrotron radiation from supernova-generated cosmic rays. As part of the LOFAR Surveys Key Science project, the Herschel-ATLAS NGP field has been surveyed with LOFAR at an effective frequency of 150 MHz. We select a sample from the MPA-JHU catalogue of SDSS galaxies in this area: the combination of Herschel, optical and mid-infrared data enable us to derive star-formation rates (SFRs) for our sources using spectral energy distribution fitting, allowing a detailed study of the low-frequency radio luminosity--star-formation relation in the nearby Universe. For those objects selected as star-forming galaxies (SFGs) using optical emission line diagnostics, we find a tight relationship between the 150 MHz radio luminosity ($L_{150}$) and SFR. Interestingly, we find that a single power-law relationship between  $L_{150}$ and SFR is not a good description of all SFGs: a broken power law model provides a better fit. This may indicate an additional mechanism for the generation of radio-emitting cosmic rays. Also, at given SFR, the radio luminosity depends on the stellar mass of the galaxy. Objects which were not classified as SFGs have higher 150-MHz radio luminosity than would be expected given their SFR, implying an important role for low-level active galactic nucleus activity.

\end{abstract}


\section{INTRODUCTION}

\label{sec:intro}

The star formation rate (SFR) of a galaxy is a fundamental parameter of its evolutionary state. Various SFR indicators of galaxies have been used in the literature over the years: for recent reviews, see \cite{2012lfr2} and \cite{2013lfr1}. In particular, two important SFR calibrations have been derived using the IR and radio continuum emission from galaxies. In the first of these, optical and ultraviolet emission from young stars (age ranges 0 -- 100 Myr with masses up to several solar masses) is partially absorbed by dust and re-emitted in the far infrared (FIR). The thermal FIR emission thus provides a probe of the energy released by star formation. On the other hand, radio emission from normal galaxies (the radio energy source is star formation, not due to accretion of matter onto a supermassive black hole, e.g. \citealt{1992lf5}) is a combination of free-free emission from gas ionised by massive stars and synchrotron emission which arises from cosmic ray electrons accelerated by supernova explosions, the end products of massive stars. Thus, radio emission (from normal galaxies) can be used as probe of the recent number of massive stars and therefore as a proxy for the SFR.

Since these processes trace star formation, one would naturally expect to see a correlation between the radio and FIR emission. \cite{1971lf2,1973lf3} showed that such a correlation exists for nearby spiral galaxies, and since then the FIR--radio correlation (FIRC, hereafter) has been the subject of many studies that have aimed to understand its physical origins and the nature of its cosmological evolution \citep[e.g.][]{1975lf7,1984lf8,1985lf1,Helou+85,1988lf6,1992lf5,2004lf40,Jarvis+10,2010lf29,Ivison+10,2011lf24,Smith+14,2015lf30,Calistro17,delhaize17}. These studies have suggested that the FIRC holds for galaxies ranging from dwarfs \citep[e.g.][]{2008lf31} to ultra-luminous infrared galaxies \citep[ULIRGs; $L_{IR}\geq10^{12.5}L_{\odot}$; e.g.][]{2001lf9} and is linear across this luminosity range. On the other hand, a number of studies \citep[e.g.][]{2003lf10,2007intr161,2008lf11} have argued that at low luminosities the FIRC may deviate from the well-known tight correlation due to the escape of the cosmic ray electrons [\cre] as a result of the small sizes of these galaxies. Although there are various factors which affect the results obtained in these studies, one contributing factor to the contradictory results might be the fact that the samples used are selected from flux-limited surveys carried out at different wavelengths (we discuss this issue in more detail in Section \ref{sec:fir-discus}).

The naive explanation of the linearity of the FIRC assumes that galaxies are {\it electron calorimeters} (all of their energy from \cre is radiated away as radio synchrotron before these electrons escape the galaxy) and {\it UV calorimeters} [galaxies are optically thick in the UV light from young stars so that the intercepted UV emission is re-radiated in the far-IR: \citep{1989lf32}]. Of these two explanations, the latter one at least is most likely incorrect, because the observed UV luminosities and the observed far-IR luminosities from SFGs are similar to each other \citep[e.g.][]{2003lf10,2005lf33}; see also \citealt{Overzier2011,Takeuchi2012,Casey2014}. This particularly breaks for low mass galaxies where the obscuration of star formation appears to be lowest \citep[e.g.][]{Bourne2012}. Furthermore, the electron calorimetry model might not hold for galaxies of Milky Way mass and below, as the typical synchrotron cooling time is expected to be longer than the inferred diffusion escape time of electrons in these galaxies \citep[e.g.][]{1996lf34}, implying that electrons may escape before they can radiate. Non-thermal radio emission has been observed in the haloes of spiral galaxies \citep[e.g.][]{2009heesen} which directly shows that the diffusion escape time of electrons is comparable to the typical energy loss time scale in some cases.

Non-calorimeter theories have also been proposed \citep{1993lf36,1997lf37,2010lf19}, often invoking a combination of processes (a `conspiracy') to explain the tightness of the FIRC. For example, \cite{2010lf19} and \cite{2010lf20} presented a non- calorimeter model taking into account different parameters (e.g. energy losses, the strength of the magnetic field and gas density etc.) as a function of the gas surface density and argued that the FIRC should break down for low surface brightness dwarfs due to the escape of \cre\ . Such models imply that stellar mass (or galaxy size) has an effect in a non-calorimeter model, as the diffusion time scale for \cre\ depends on the size of a galaxy.

To date the radio luminosity--SFR relation and the FIRC have been studied almost exclusively at GHz bands \citep[e.g.][]{2001lf9,davies+17}, because sensitive radio surveys have mostly been carried out at these radio frequencies \citep[e.g.][]{1995ref118,1998ref117}. Due to the lack of available data, in most previous work the radio luminosity of SFGs has been considered as a function of SFR only. However, there is a well-known tight relation (the `main sequence' of star formation) which has been observed between SFR and stellar mass of SFGs with a $\sim$0.3 dex scatter \citep[e.g.][]{2007lf48}. This relation holds for SFGs in the local Universe \citep[e.g.][]{2004ref8,2007lf49} and most likely evolves with redshift \citep[e.g.][]{2011lf50,Johnston2015}. This tight relation gives an additional argument that the mass or size of the host galaxy should be taken into account when considering the radio luminosity/SFR relation.

 With new radio interferometer arrays such as the Low Frequency Array \citep[LOFAR;][]{vanHaarlem2013}, we are able to move toward lower radio frequencies, where the contribution to the radio luminosity from thermal free-free emission becomes increasingly negligible although synchrotron self absorption might become more important \citep[e.g.][]{israel92,kapinska17,schober+17}. In addition, with the increasing number of surveys at other wavebands, it is possible to use multiwavelength data sets to derive galaxy properties (such as SFR, galaxy mass etc.) using spectral energy distribution (SED) modelling. Recently, \cite{Calistro17} investigated the IR-radio correlation of radio selected SF galaxies over the Bo\"otes field \citep{2016williams} using LOFAR observations and SED fitting and were able to show that SFGs show spectral flattening towards low radio frequencies (probably due to environmental effects and ISM processes).

 The goal of the present paper is to investigate the relationship between low-frequency radio luminosity, using LOFAR observations at 150 MHz over the Herschel Astrophysical Terahertz Large Area Survey (H-ATLAS) North Galactic Pole (NGP) field ($\sim$142 square degrees), and the physical properties of galaxies such as SFR and stellar mass, using multiwavelength observations available over the field. The results obtained in this work will be crucial for the interpretation of future surveys.

 The layout of this paper is as follows. A description of the sample, classification and data are given in Section 2. Our key results are given in Section 3, where we present the results of our regression analysis using Markov-Chain Monte Carlo methods (MCMC) and stacking. In Section 4 we interpret our findings and summarize our work.  Our conclusions are given in Section 5.

Throughout the paper we use a concordance cosmology with $H_0=70$ km s$^{-1}$ Mpc$^{-1}$, $\Omega_{m}=0.3$ and $\Omega_{\Lambda}=0.7$. Spectral index $\alpha$ is defined in the sense $S \propto \nu^{-\alpha}$.

\section{DATA}

\subsection{Sample and emission-line classification}

 To construct our sample we selected galaxies from the seventh data release of the Sloan Digital Sky Survey \citep[SDSS DR7; ][]{2009ref113} catalogue with the value-added spectroscopic measurements produced by the group from the Max Planck Institute for Astrophysics, and the John Hopkins University (MPA-JHU)\footnote{\url{http://www.mpa- garching.mpg.de/SDSS/}} in the H-ATLAS \citep{2010ref1} NGP field. This provided a parent sample of 16,943 SDSS galaxies over the HATLAS/NGP field. Since the radio maps do not fully cover the H-ATLAS/NGP field only, 15,088 sources (out of 16,943 galaxies) have a measured LOFAR flux density, spanning the redshift range $0<z<0.6$. The sample does not include quasars because they outshine the host galaxies for these objects which makes it difficult to study the host galaxy properties.

\citet*[][BH12 hereafter]{2012wref29} have constructed a radio-loud active galactic nuclei (AGN) sample by combining the MPA-JHU sample with the National Radio Astronomy Observatory (NRAO) Very Large Array (VLA) Sky Survey \citep[NVSS;][]{1998ref117} and the Faint Images of the Radio Sky at Twenty centimetres (FIRST) survey \citep{1995ref118} following the methods described by \cite{2005ref57} and \cite{2009ref169}. Here we briefly summarize their methods: further details are given by the cited authors. First, each SDSS source was checked to see whether it has an NVSS counterpart: in the case of multiple-NVSS-component matches the integrated flux densities were summed to obtain the flux density of a radio source. If there was a single NVSS match, then the FIRST counterparts of the source were checked. If a single FIRST component was matched, the source was accepted or rejected based on the source's FIRST flux. If there were multiple FIRST components the source was accepted or rejected based on its NVSS flux.

We firstly cross-matched the MPA-JHU sample with the BH12 catalogue in order to construct our radio AGN sub-sample (with 279 members). Some of these radio sources have emission-line classifications (i.e. they were classified by BH12 as high-excitation radio galaxies, HERGs, or low-excitation radio galaxies, LERGs). A number of radio sources have no clear emission-line classification and these are shown as HERG/LERG? (with 86 objects) in the corresponding tables and figures. The remaining galaxies from the 15,088 sources were classified as SFGs (with 4157 sources), Composite objects (with 1179 objects), Seyferts (with 328 objects), LINERs (with 117 members) and Ambiguous sources [341 members; these are objects that are classified as one type in the [NII/H$_{\alpha}$]-diagram and another type in the [SII/H$_{\alpha}$]-diagram using the modified BPT-type \citep{1981ref115} emission-line diagnostics described by \cite{2006ref2}]. This classification was carried out only using the following emission lines: [NII] $\lambda$6584, [SII] $\lambda$6717, H$_{\beta}$, O$_{\mathrm{III}}$ $\lambda$5007 and H$_{\alpha}$. Composite objects were separated from star-forming objects using the criterion given by \cite{2003ref88}. It is necessary for objects to have the required optical emission lines -- in our case H$_{\beta}$, O$_{\mathrm{III}}$ $\lambda$5007, H$_{\alpha}$, [NII] $\lambda$6584 and [SII] $\lambda$6717) -- detected at $3\sigma$ in order to classify galaxies accurately. This requirement limits the classification of SFGs to around $z\leq0.25$: biases due to this selection are discussed in section \ref{sec:lofar-flux}. As we move to higher redshifts ($z\geq0.25$) we cannot detect strong optical emission lines from normal star-forming objects. Only a few high-redshift SFGs ($z>0.25$) have optical emission lines detected at 3$\sigma$ and these galaxies are probably starbursts at higher redshifts. A large number of galaxies, more than half the parent sample, are not detected in all the required optical emission lines at 3$\sigma$, and those are therefore unclassified by these methods (with 8687 members). In order to make a direct comparison we include only sources with $z\le 0.25$ in figures in which we compare the different classes of galaxies.

\begin{table*}

\caption{The number of sources in the whole sample and in each
  population after optical emission-line classification, together with
  their $3\sigma$ detection rate in FIR, radio and both
  wavebands. \label{sample-table}}
\begin{tabular}{lrrrrrrr}
\hline
Population type & Classified & Herschel 3$\sigma$ & LOFAR 3$\sigma$ & Detected in both bands & \magphys & Average stellar mass & Average SFR \\
&&&&&good fit counts&$(M_{\odot}$)&($M_{\odot}$ yr$^{-1}$)\\
\hline
SFGs & 4157 & 3393 & 2369 & 2179 & 3908 & 1.71e+10 & 2.21 \\
Composites & 1179 & 980 & 739 & 677 & 1133 & 6.10e+10 & 2.22 \\
Seyferts & 328 & 205 & 201 & 148 & 274 & 7.74e+10 & 0.91 \\
RL AGN (HERGs/LERGs) & 193 & 26 & 190 & 26 & 172 & 3.10e+11 & 0.88 \\
HERG/LERG? & 86 & 5 & 86 & 5 & 75 & 3.61e+11 & 2.19 \\
LINERs & 117 & 59 & 49 & 33 & 109 & 9.12e+10 & 0.26 \\
Unclassified by BPT & 8687 & 2184 & 2880 & 1064 & 8029 & 1.54e+11 & 0.54 \\
Ambiguous & 341 & 228 & 197 & 163 & 317 & 8.81e+10 & 1.44 \\
\hline
\end{tabular}
\end{table*}

\subsection{Radio data}

\subsubsection{Flux densities at 150 MHz}

\label{sec:lofar-flux}

LOFAR observed the H-ATLAS NGP field as one of several well-studied fields observed at the sensitivity and resolution of the planned LOFAR Two-Metre Sky Survey \citep{Shimwell+16}. The observations and calibration are described by \citet*[][hereafter H16]{Hardcastle+16}, but for this paper we use a new direction-dependent calibration procedure. This processing of the H-ATLAS data will be described in more detail elsewhere but, to summarize briefly, it involves replacing the facet calibration method described by H16 with a direction-dependent calibration using the methods of \cite{Tasse14b,Tasse14a}, implemented in the software package {\sc killms}, followed by imaging with a newly developed imager {\sc ddfacet} \citep{Tasse17} which is capable of applying these direction-dependent calibrations in the process of imaging. The H-ATLAS data were processed using the December 2017 version of the pipeline, {\sc ddf-pipeline}\footnote{See \url{http://github.com/mhardcastle/ddf-pipeline} for the code.}, that is under development for the processing of the LoTSS survey \citep[][and in prep.]{Shimwell+16}. The main advantage of this reprocessing is that it gives lower noise and higher image fidelity than the process described by H16, increasing the point-source sensitivity and removing artefacts from the data, but it also allows us to image at a slightly higher resolution -- the images used in this paper have a 6-arcsec restoring beam.

Radio flux densities at 150 MHz for all the SDSS galaxies in our sample (15088 sources) were directly measured from the final full- bandwidth LOFAR maps. We took the flux extracted from the image in an aperture of 10 arcsec in radius for all MPA-JHU galaxies at the SDSS source positions, which was chosen considering the resolution of the LOFAR maps. With this extraction radius, the aperture correction is negligible. The noise-based uncertainties on these flux densities were estimated using the LOFAR r.m.s. maps: we discuss the flux scale and the checks we carried out on the forced-photometry method in Appendix \ref{app:fluxscale}. To convert the 150-MHz flux densities to 150-MHz luminosities ($L_{150}$ in W Hz$^{-1}$) we adopt a spectral index $\alpha=0.7$ [the typical value that was found by H16]. As mentioned above the radio maps do not fully cover the H-ATLAS/NGP field: only 15,088 sources have a measured LOFAR flux density of which $\sim 50$ per cent were detected at the $3\sigma$ level. Counts and detection statistics of the whole sample with LOFAR flux measurements are given in Table \ref{sample-table}. We consider only those sources with LOFAR flux density measurements (including non-detections) from now on.

\begin{figure*}

\begin{center}

\scalebox{1.5}{

\begin{tabular}{cc}

\includegraphics[width=11.5cm,height=11.5cm,angle=0,keepaspectratio]{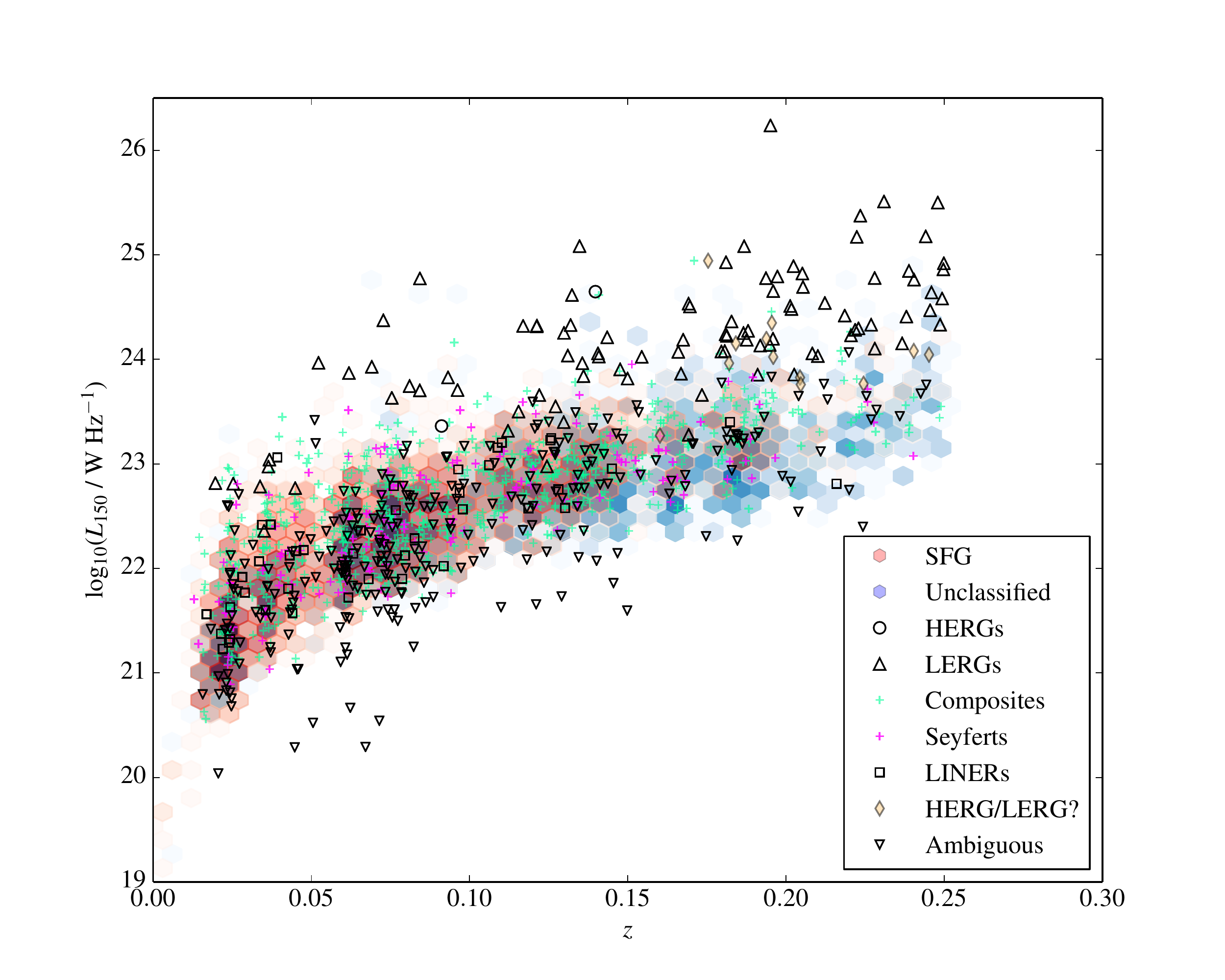}\\

\end{tabular}}

\caption{Combined density/scatter figure showing the distribution of the
  luminosity at 150 MHz of all LOFAR-detected galaxies in the sample
  as a function of their redshifts. Colours and different symbols
  indicate different emission-line classes. Hexagons show the density of
  the two most populated classes, SFG (salmon) and objects unclassified
  on a BPT diagram (blue). Other sources are shown with a symbol for
  each object: these include composites (light green crosses), Seyferts (magenta crosses) and LINERs (open black squares) and Ambiguous sources (open triangle point down). Radio AGN classified by their
  emission lines are also shown: HERGs (open black circles) and LERGs
  (open black triangles). Sources in the BH12 sample that were not
  classified either as a HERG or LERG are shown as light yellow diamonds. It is worth noting that an 
overlap is seen between RL AGN (HERGS and LERGS) and emission line 
classification (Seyferts/LINERS). Similar results were also reported by \citet{smolcic+9}. \label{lum}}
\end{center}
\end{figure*}

Fig. \ref{lum} shows the 150-MHz luminosity distribution of the detected galaxies as a function of redshift. H16 showed that the radio luminosity function (at 150 MHz) of SFGs selected in the radio shows an evolution with redshift (within $0.0<z<0.3$) which they suggest is a result of the known evolution of the star formation rate density of the Universe over this redshift range. As can be seen from Fig. \ref{lum}, we include all SFGs in the sample with a similar redshift range ($0.0<z<0.3$), because this allows us to investigate any variation in the relations studied here for the star-forming populations with different luminosities at relatively low redshifts. We take into account possible degeneracies between redshift and luminosity when we interpret our results, but a priori we do not expect any particular change in the physics of individual galaxies over this redshift range, and it is that which drives the radio--SFR and radio--FIR correlations.

\subsubsection{Flux densities at 1.4 GHz with FIRST}

\label{sec:spix}

We obtained the FIRST \citep{1995ref118} images and r.m.s. maps of the H-ATLAS/NGP field and, as for the LOFAR flux density measurements, we measured the flux densities at the source positions, also within an aperture 10 arcsec in radius. Uncertainties on these flux densities were estimated in the same way as for the LOFAR flux errors, using the 1.4-GHz r.m.s. maps. The 1.4-GHz luminosities of the sources in the sample ($L_{1.4}$) were estimated using these flux densities and a spectral index $\alpha = 0.7$ at the spectroscopic redshift. To check the relative flux scales, we estimated the spectral index for each of the 2930 sources detected at the $3\sigma$ level in both FIRST and LOFAR data. We emphasize that this is not a true estimate of the population spectral index, and the biases are complex because the LOFAR data are deeper than FIRST in some areas of the sky and shallower in others. However, a simple flux scale check and a comparison of the populations are possible. The median spectral indices and their $1\sigma$ errors from bootstrapping, along with the median LOFAR flux density, are given in Table \ref{spix-table}. We see that the median spectral index, and the spectral indices of most individual populations, are close to the value of 0.7 that we assume in calculating the luminosity. There is in general no way of determining a source's emission-line type from its two-point spectral index. Interestingly, though, the sources unclassified by emission-line diagnostics seem to have significantly flatter spectra than the others. This may be a selection bias of some kind, given that they also tend to be at higher redshift, or it may indicate some physical difference in the origin of their emission. We return to this point in section \ref{sec:unclassified}.

\cite{Sargent+10} carried out a survival analysis and showed that selecting samples from flux limited surveys can introduce a selection bias which would eventually lead to misleading results. For this reason, we further carried out a doubly-censored survival analysis in order to calculate the median spectral index ($\alpha^{1.4GHz}_{150MHz}$) for each population. This is necessary because the spectral indices will have both upper and lower limits due to non-detections in either wavebands. Below we briefly explain the survival analysis tool\footnote{This tool is written in Perl/PDL by M.\ Sargent (private communication). The Perl Data Language (PDL) has been developed by K. Glazebrook, J. Brinchmann, J. Cerney, C. DeForest, D. Hunt, T. Jenness, T. Luka, R. Schwebel, and C. Soeller and can be obtained from http://pdl.perl.org.} we used for this work and we refer the reader to the source paper presented by \cite{Sargent+10}. The doubly-censored survival analysis tool that was used here uses the method described by \cite{Schmitt+93}, which requires no assumptions about the form of the true distribution of $\alpha$. The method redistributes the upper and lower limits in order to derive a doubly-censored distribution function. Median estimates of $\alpha^{1.4GHz}_{150MHz}$, with their errors derived from the survival analysis are also given in Table \ref{spix-table}. A comparison of the spectral indices obtained using the survival analysis with the medians using only detected sources shows that taking into account left- and right-censored data leads to steeper spectral indices, which are much closer to 0.7.

\begin{table*}

  \caption{Spectral index between 150 MHz and 1.4 GHz for the sources
    detected by both LOFAR and FIRST.}
  \label{spix-table}
  \begin{center}
  \begin{tabular}{lrrrr}
    \hline
    Category&Number&Median $\alpha$&Median 150-MHz &Median $\alpha$\\
    &&&flux density (mJy)&(Survival analysis results)\\
    \hline
All & 3073 & $0.47 \pm 0.01$ & $2.94 \pm 0.10$ & 0.53$_{-0.06}^{+0.03}$ \\
SFG & 1089 & $0.49 \pm 0.01$ & $2.58 \pm 0.08$ & 0.58$_{-0.07}^{+0.03}$\\
Unclassified & 1106 & $0.37 \pm 0.02$ & $2.13 \pm 0.11$ & 0.42$_{-0.07}^{+0.06}$\\
Radio-loud & 274 & $0.60 \pm 0.01$ & $25.69 \pm 2.97$ & $0.60 \pm 0.01$\\
Composite & 374 & $0.52 \pm 0.03$ & $3.36 \pm 0.22$ & 0.61$_{-0.05}^{+0.06}$\\
Seyfert & 110 & $0.49 \pm 0.04$ & $3.38 \pm 0.34$ & 0.56$_{-0.09}^{+0.08}$\\
LINER & 26 & $0.47 \pm 0.08$ & $3.13 \pm 0.79$ & 0.47$_{-0.17}^{+0.30}$\\
Ambiguous & 94 & $0.51 \pm 0.05$ & $3.37 \pm 0.23$ & 0.57$_{-0.08}^{+0.13}$\\
\hline
  \end{tabular}
  \end{center}
\end{table*}

\subsection{Far-IR data}

\label{sec:firdata}

$\textit{Herschel}$-ATLAS  provides imaging data for the $\sim$ 142 square degrees NGP field using the Photo-detector Array Camera and Spectrometer \cite[PACS at 100 and 160 $\mu$m: ][]{2010ref10,poglitsch10} and the Spectral and Photometric Imaging Receiver \citep[SPIRE at 250, 350 and 500 $\mu$m:][]{2010ref11,2011ref145,2016valiante}. To derive a maximum- likelihood estimate of the flux densities at the positions of objects in the SPIRE bands whether formally detected or not, the point spread function (PSF)-convolved H-ATLAS images were used for each source together with the errors on the fluxes. Further details of the flux measurement method are given by \cite{2010ref7,2013ref9}.

In order to estimate 250-$\mu$m luminosities ($L_{250}$ in W Hz$^{-1}$) for our sources we assumed a modified black-body spectrum for the far-IR SED (using both SPIRE and PACS bands); we fixed the emissivity index $\beta$ to 1.8 [the best-fitting value derived by \cite{2013ref9} and \cite{2013ref87} for sources in the H-ATLAS at these redshifts] and obtained the best fitting temperatures, integrated luminosities ($L_{\mathrm{IR}}$) and rest-frame luminosities at 250 $\mu$m ($L_{250}$) by minimizing $\chi^2$ for all sources with significant detections. To calculate the 250-$\mu$m $k$-corrections the same emissivity index and the mean of the best-fitting temperatures for each emission-line class were then used. These corrections were included in the derivation of the 250-$\mu$m luminosities that are used in the remainder of the paper. $k$-corrections are, naturally, small for our sample because of the low maximum redshift of our targets.

 Stacked measurements in confused images can be biased by the presence of correlated sources because the large PSF can include flux from nearby sources. Several methods have been proposed to account for this bias, including the flux measurements in GAMA apertures by \cite{2012ref168}. This method explicitly deblends confused sources and divides the blended flux between them using PSF information. However, we checked for the effects of clustering in our previous work in which we used the same sample and sample classification \citep{2015lf17}. The results of this analysis indicated that our work is not biased by the effects of clustering and that the results are robust.

\subsection{Star formation rates (SFRs)}

 Due to the large range of multi-wavelength data available over the \hatlas\ NGP field, we can model the properties of the sources that we observe consistently, using all of the photometric data simultaneously. One way of doing this is with the widely-used \magphys\ code \citep{dacunha08}. At the heart of \magphys\ is the idea that the energy absorbed by dust at UV-optical wavelengths is re-radiated in the far-infrared. This `energy balance' thus forces the entire SED from UV to millimetre wavelengths to be physically consistent, providing a greater understanding of a galaxy's properties than would be obtained by studying either the starlight or dust emission in isolation.

 The precise details of the \magphys\ model are discussed in detail by \citet{dacunha08}, but the main components of the model can be summarized as follows. \magphys\ comes with two libraries; the first contains stellar model SEDs, while the second includes dust models. The stellar library we use is based on the latest version of the \citet{bc03} simple stellar population library (often referred to as the CB07 models, unpublished). Exponentially declining star formation histories are assumed, with stochastic bursts of star formation superposed, such that approximately half of the star formation histories in the library have experienced a burst in the last 2 Gyr. These stellar models are then subjected to the effects of a two-component dust model from \citet{charlot00}, in which the two components correspond to the stellar birth clouds (which affects only the youngest stars) and the ambient interstellar medium (ISM).

Each dust SED in the \magphys\ \footnote{\magphys\ uses the initial mass function (IMF) of \cite{Chabrier03}.} library consists of multiple optically thin modified blackbody profiles \citep[e.g.][]{hildebrand83,hayward12,2013ref87}, with variable normalization, temperature and emissivity indices to describe dust components of different sizes. For example, stellar birth clouds are modelled with $\beta_{\mathrm{BC}} = 1.5$ and temperature $30 < T_{BC}< 60$\,K, while the ambient ISM is modelled with $\beta_{\mathrm{ISM}}= 2.0$ and $15 < T_{ISM} < 25$\,K. Also included is a model for emission from polycyclic aromatic hydrocarbons, which are readily apparent in the mid-infrared.

Of particular importance for this study is the fact that \magphys\ has recently been shown to recover the properties of galaxies (e.g. stellar mass, star formation rate, dust mass\slash luminosity) reliably, irrespective of viewing angle, evolutionary stage, and star formation history \citep{hayward15}. Though the current version of \magphys\ does not include any AGN emission in the modelling, \cite{hayward15} also showed that acceptable fits and reliable parameters could be recovered even in the case where a merger-induced burst of AGN activity is producing up to 25 per cent of a source's total bolometric luminosity. \magphys\ has been extensively used both within the \hatlas\ survey and elsewhere in the literature \citep[e.g.][and many more]{dacunha10,smith12,berta13,lanz13,brown14,negrello14,rowlands14,eales15,smith15,dariush16}.

 In this work, we use the precise spectroscopic redshifts of the SDSS sample along with \magphys\ to model the 14 bands of photometric data available over the \hatlas\ NGP field consistently. These bands include the SDSS $ugriz$ bands \citep{york00}, data from the {\it WISE} survey \citep{wright10} in bands centred on 3.4, 4.6, 12 and 22\,$\mu$m, and data from the \hatlas\ survey from the PACS (centred on 100 and 160\,$\mu$m) and SPIRE (centred on 250, 350 and 500\,$\mu$m) measured as discussed in the previous section. The \magphys\ model libraries are redshifted, and passed through the filter curves for each of these bands, before those combinations of stellar and dust components which satisfy the energy balance criterion are included in the fitting, estimating the $\chi^2$ goodness-of-fit parameter for every valid combination, allowing the best-fitting model and parameters to be identified. Assuming that $P \equiv \exp \left(-\chi^2/2\right)$, it is then possible to derive marginalized probability distribution functions (PDFs) for every parameter in the model. We can also derive standard uncertainties on each parameter derived according to half of the interval between the 16$^{\mathrm{th}}$ and 84$^{\mathrm{th}}$ percentiles of the PDF.

To ensure that the measured fluxes in different bands are as consistent as possible, we define our input photometry as follows. As recommended by the SDSS documentation,\footnote{The relevant SDSS documentation can be found at \url{http://www.sdss.org/dr12/algorithms/magnitudes/}} we use the SDSS MODEL magnitudes to estimate the most precise colours, and `correct-to-total' using the difference between the cMODEL and MODEL magnitudes in the $r$ band. We apply the 0.04 and 0.02 magnitude corrections in the $u$ and $z$ bands, respectively, recommended by \citet{bohlin01} to convert from the SDSS photometric system to the AB system \citep{oke83}. Finally, we correct the SDSS magnitudes for Galactic extinction using the SDSS extinction values computed at the position of each object using the prescription of \citet{schlegel98}.

At mid-infrared wavelengths, we adopt photometry from the UN{\it WISE} \citep{lang14a} reprocessing of the {\it WISE} images, which uses the method of \citet{lang14b} to perform forced photometry on unblurred co-adds of the {\it WISE} imaging, using shape information from the SDSS $r$ band to ensure consistency with the SDSS magnitudes. As for the {\it Herschel} data (see below), we include the {\it WISE} photometry in the fitting even when objects are formally not detected (i.e., if a source has $<3\sigma$ significance in a particular band pass). \magphys\ is the ideal tool for dealing with this type of data; since it is based on $\chi^2$ minimization, the low-significance data points can naturally be treated consistently with the other wavelengths, and it is unnecessary to consider applying e.g. `upper limits' (which can introduce discontinuities in the derived PDFs, and essentially ignore information below some arbitrary threshold). Furthermore, \citet{2013ref87} demonstrated that formally non-detected photometry in the PACS bands is useful when it comes to deriving robust effective dust temperatures, which are biased in their absence.

 To account for residual uncertainties in the aperture definitions and zeropoint calibrations, we add 10 percent in quadrature to the SDSS, {\it WISE} and {\it PACS} photometry, and 7 percent in quadrature to the {\it SPIRE} photometry, following \citet{smith12}. We identify bad fits using the method of \citet[][see their appendix B]{smith12}, who used a suite of realistic simulations to define a limiting $\chi^2$ value as a function of the number of bands of photometry available, above which there is less than one per cent chance that the photometry is consistent with the model. It is worth noting that sources with bad \magphys\ fits (with 1071 objects) are not included in any of the analyses carried out here. We also exclude these sources from all of the figures presented in this paper.

\begin{figure*}

\begin{center}

\scalebox{1.2}{

\begin{tabular}{c}

\includegraphics[width=11cm,height=11cm,angle=0,keepaspectratio]{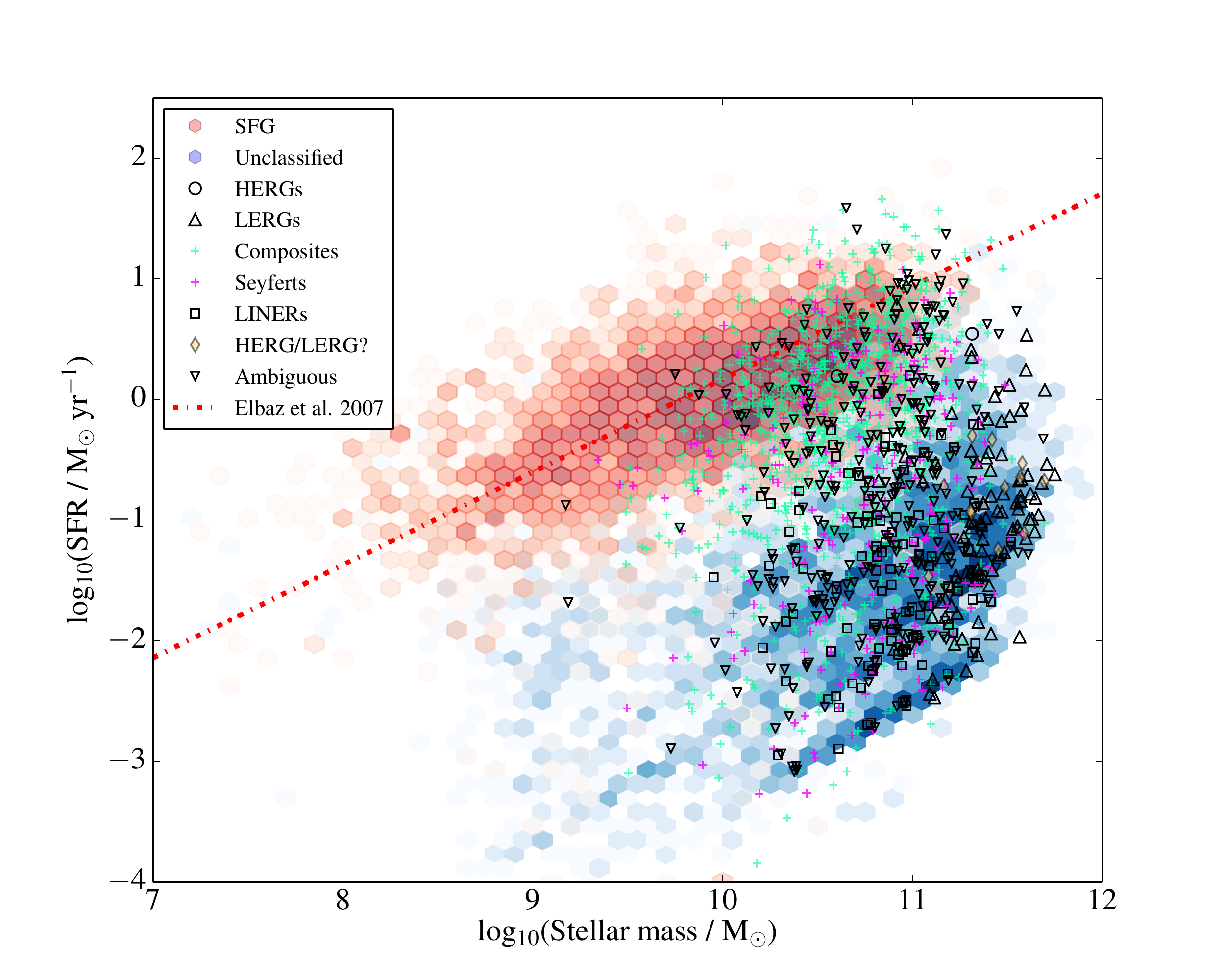}\\

\includegraphics[width=11cm,height=11cm,angle=0,keepaspectratio]{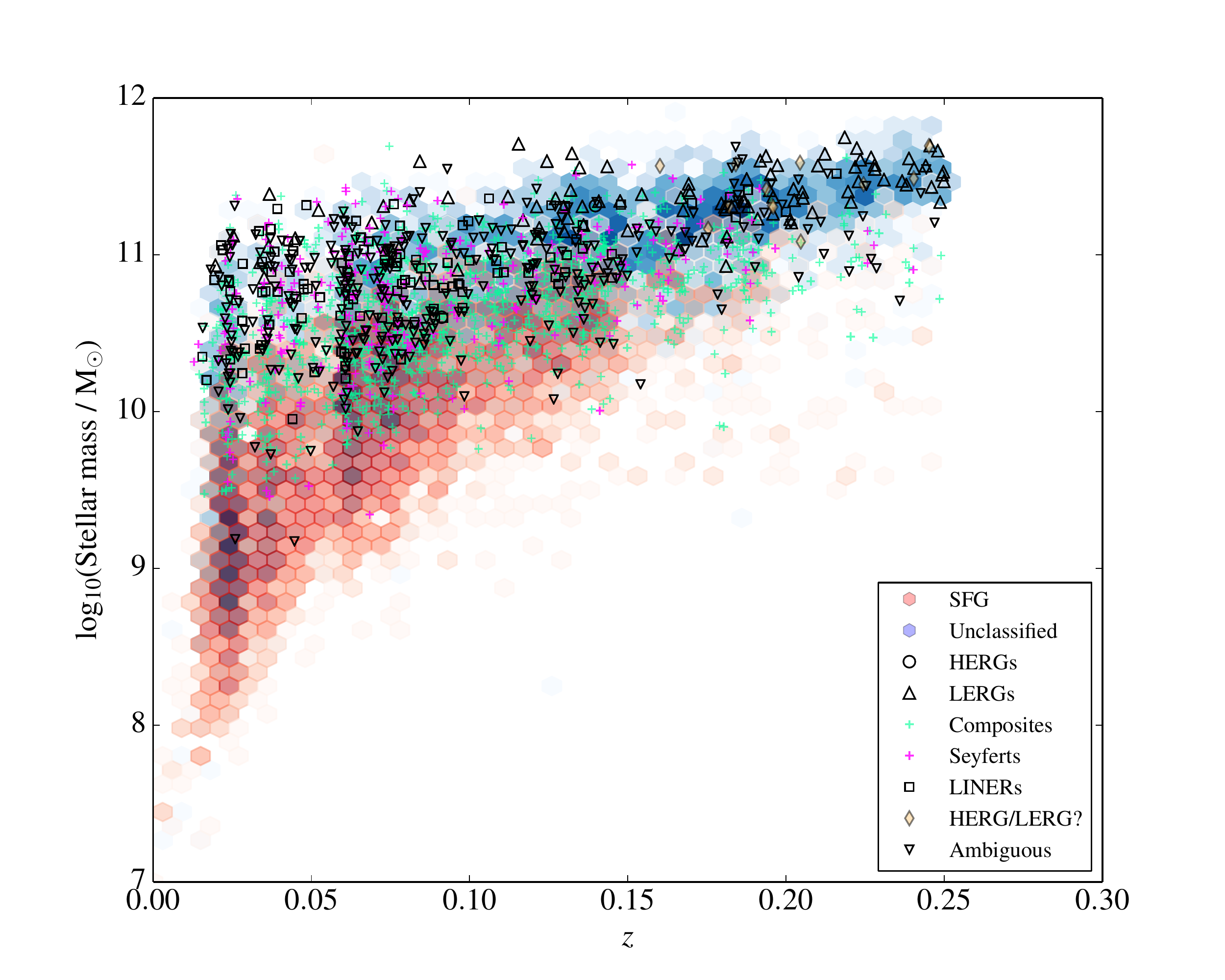}\\

\end{tabular}}

\caption{Top panel: The distribution of best-fitting \magphys\ SFRs of all galaxies in the sample as a function of their \magphys\ stellar masses. The dash-dot line shows the local main sequence relation \citep{Elbaz07}. Diagonal lines in the lower part of this panel are the result of discreteness in the specific SFR of the models fitted in \magphys: this does not affect results in the paper as the SFR of non-SFG are not used in any quantitative analysis. Bottom panel: \magphys\ stellar mass distribution of all galaxies in the sample versus redshift. Symbols and colours as for Fig. \ref{lum}. \label{mass-sfr}}
\end{center}
\end{figure*}

 The key results of the \magphys\ modelling are shown in Fig.\ \ref {mass-sfr}, where we show the inferred star-formation rate against stellar mass for the whole sample, colour-coded by their emission-line classification. We see a clear `main sequence' of star formation inhabited by objects whose emission lines classify them as star-forming galaxies. Objects unclassified on the emission-line diagram tend on the whole to lie off the `main sequence', with low SFR for their mass; many of these are likely to be passive (quiescent) galaxies. Some composites and Seyferts lie on the main sequence, most LINERs and objects classed by BH12 as radio galaxies lie below it, but in all cases a minority of objects do not follow the trend. The bottom panel of Fig.\ \ref{mass-sfr} shows that the ability to classify with emission-line diagnostics is essentially lost by $z \sim0.25$, but most galaxies in the sample above this redshift are massive ($\sim 10^{11.5} M_\odot$). Therefore, these are most likely passive galaxies that have moved away from the star formation main sequence: the fact that most hosts of BH12 radio galaxies lie in this region of the figure is consistent with this interpretation.

 Here, and throughout the paper, we make use of the \magphys\ best-fitting SFRs and masses rather than the Bayesian estimates. This is because a number of the objects in the whole sample have low SFRs and the prior used in \magphys\ is effectively biased towards higher sSFR (specific star formation rate), in the sense that there are more templates with higher sSFR values. The quoted uncertainties on the best-fitting values are half of the interval between the 16th and 84th percentiles of the PDF.

 We note finally that the \magphys\ SFRs and galaxy masses are generally very similar to those already provided in the MPA-JHU catalogue ($\sim 0.2$dex difference), suggesting that there are no very serious biases in our analysis. The advantages of using \magphys\ is that we can incorporate the H-ATLAS and WISE data available for this sample in a consistent way and that we are unlikely to be affected by reddening. In Appendix \ref{app:magphys} we compare \magphys\ SFRs to H${\alpha}$-SFR and discuss in more detail why the \magphys\ SFR estimates were used in this work.

\section{Analysis and results}

\subsection{Radio luminosity and star formation}

\begin{figure*}

\scalebox{1.3}{

\begin{tabular}{c}

\includegraphics[width=11cm,height=11cm,angle=0,keepaspectratio]{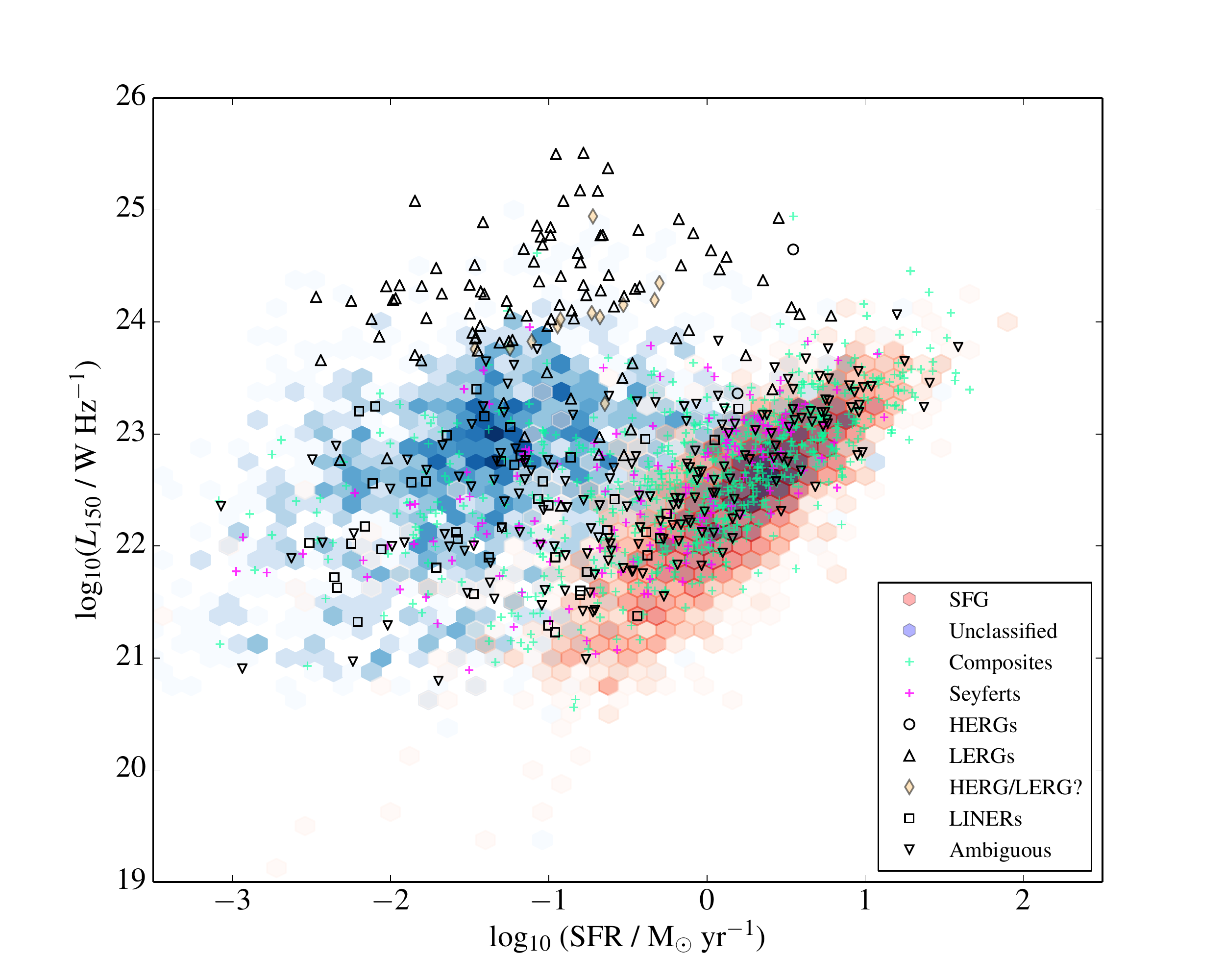}\\

\includegraphics[width=11cm,height=11cm,angle=0,keepaspectratio]{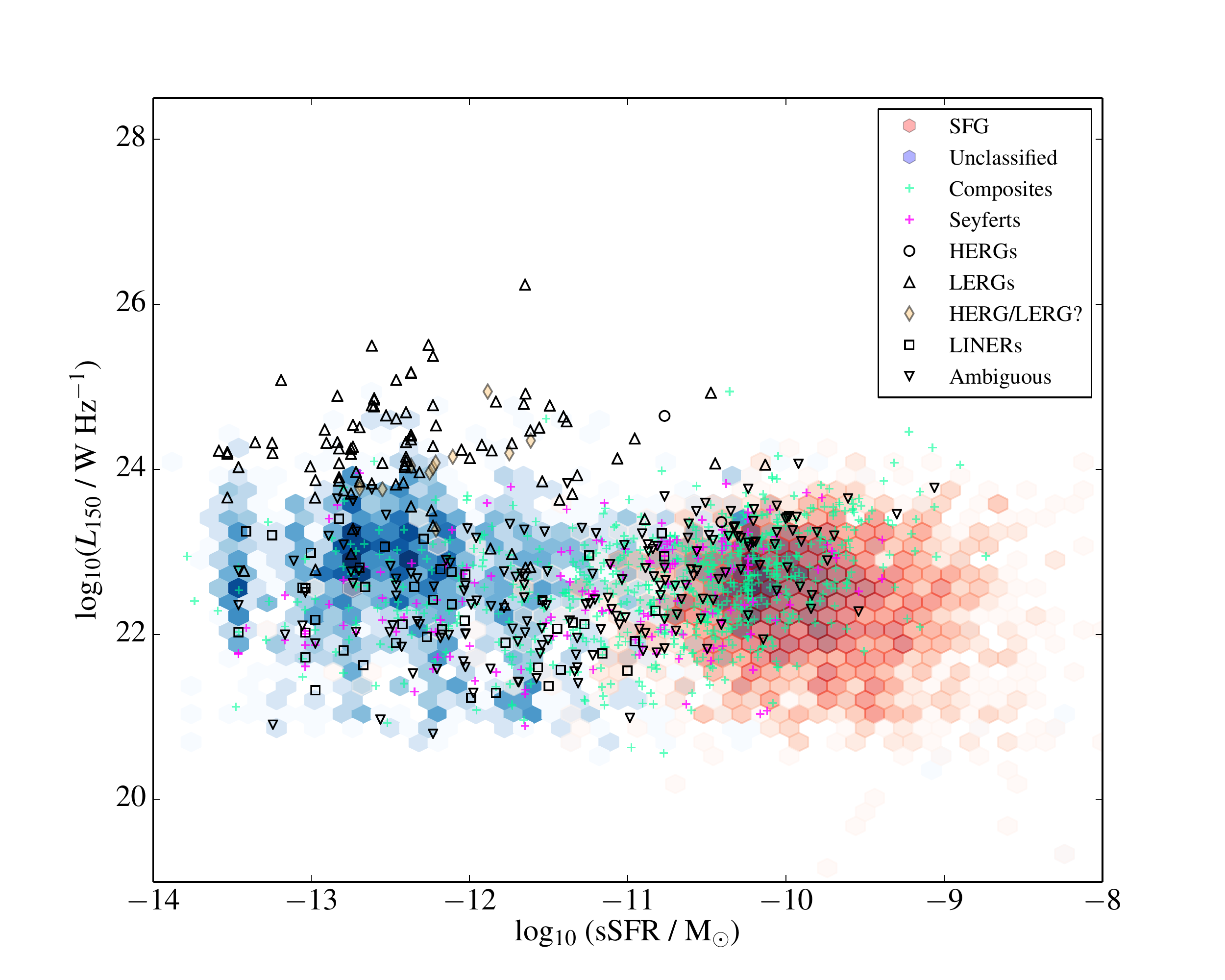}\\

\end{tabular}}
\caption{Top panel: The distribution of LOFAR 150-MHz luminosities of different classes of objects detected by LOFAR at the $3\sigma$ level as a function of their SFRs. Bottom panel: The distribution of LOFAR 150-MHz luminosities of different classes of objects detected by LOFAR at the $3\sigma$ level as a function of their sSFRs. Symbols and colours as for Fig. \ref{lum}. \label{lfr-sfr-all} }
\end{figure*}

 In Fig. \ref{lfr-sfr-all} we show the distribution of $L_{150}$ of all classified galaxies in the sample as a function of their best-fitting \magphys\ SFRs. It should be noted that objects undetected by LOFAR are not plotted to provide a clear presentation. There are several interesting features of this figure.  Firstly, we see a clear correlation between SFR and $L_{150}$ for the star-forming objects in the bottom right of the figure: this is the expected $L_{150}$-SFR relation which we will discuss in the following section. Known RLAGN from the BH12 catalogue occupy the top left part of the diagram, as expected because radio emission from AGN will be much higher than for normal galaxies. However, a very large fraction of the sources unclassified on the basis of emission lines (which are, as shown above, mostly massive galaxies lying off the main sequence of star formation) lie above the region occupied by SFGs. These sources clearly have higher radio luminosities than normal galaxies with the same SFR but tend to be less radio luminous than the BH12 radio AGN. Some unclassified objects lie in the SFG locus, and some SFGs lie on the locus of unclassified objects, but on the whole the two populations are strikingly distinct on this figure. Unclassified sources by BPT diagrams were also found as distinct population by \cite{leslie16} who studied the relation between star formation rate and stellar mass of a sample selected from the MPA-JHU. We discuss the nature of the unclassified objects in Section \ref{sec:unclassified}. Objects classed as Seyferts and composites mostly lie in the upper part of the region occupied by SFGs, where we see objects with higher radio luminosities and SFRs though with a scatter up towards the RLAGN in some cases.

\subsection{The low-frequency radio luminosity--SFR relation}

\label{sec:rsfr}

\begin{table*}

\caption{Results of stacking $L_{150}$ of SFGs in bins of SFR. Column 1: The range of stellar mass spanned by the bin. Column 2: The range of SFR spanned by the bin. Column 3: The mean redshift in each SFR bin. Column 4: The total number of sources in each bin. Column 5: The calculated weighted average of $L_{150}$. Column 6: Ratio of the mean of $L_{150}$ to the mean of SFR in each SFR bin. Column 7: Results of the KS test that was carried out to evaluate whether sources in the bins were significantly detected. \label{tbl-lfr-sfr0}}
\begin{tabular}{lrrrrrr}
\hline
Stellar mass range &SFR range  & Mean $z$ & $N$ & $L_{150}$ &$L_{150}$/$SFR$& KS probability \\
($M_{\odot}$)&($M_{\odot}$ yr$^{-1}$)&&&($\times 10^{22}$ W Hz$^{-1}$)&($\times
10^{22}$ W Hz$^{-1}$ $M_\odot^{-1}$ yr)\\
\hline
$6.\leq \log_{10}(M_{*})<9.5$ & 0.001 -- 0.01 & 0.04 & 16 & 0.06$^{+0.03}_{-0.04}$ & 23.44$^{+21.68}_{-21.68}$ & $\ll 1\%$ \\
 & 0.01 -- 0.03 & 0.04 & 40 & 0.01$^{+0.01}_{-0.01}$ & 0.24$^{+0.50}_{-0.50}$ & 0.18 \\
 & 0.03 -- 0.1 & 0.04 & 138 & 0.03$^{+0.01}_{-0.01}$ & 0.44$^{+0.14}_{-0.14}$ & $\ll 1\%$ \\
 & 0.1 -- 0.3 & 0.04 & 408 & 0.08$^{+0.01}_{-0.01}$ & 0.47$^{+0.07}_{-0.07}$ & $\ll 1\%$ \\
 & 0.3 -- 1 & 0.05 & 387 & 0.23$^{+0.03}_{-0.03}$ & 0.47$^{+0.06}_{-0.06}$ & $\ll 1\%$ \\
 & 1 -- 3 & 0.07 & 169 & 0.81$^{+0.08}_{-0.09}$ & 0.60$^{+0.06}_{-0.06}$ & $\ll 1\%$ \\
 & 3 -- 10 & 0.08 & 29 & 3.26$^{+0.77}_{-0.73}$ & 0.78$^{+0.18}_{-0.18}$ & $\ll 1\%$ \\
$9.5\leq \log_{10}(M_{*})<13.$ & 0.001 -- 0.01 & 0.07 & 16 & 0.36$^{+0.13}_{-0.12}$ & 81.66$^{+36.94}_{-36.94}$ & $\ll 1\%$ \\
 & 0.01 -- 0.03 & 0.06 & 20 & 0.46$^{+0.37}_{-0.34}$ & 24.86$^{+19.34}_{-19.34}$ & 0.09 \\
 & 0.03 -- 0.1 & 0.06 & 32 & 0.73$^{+0.41}_{-0.48}$ & 12.34$^{+7.53}_{-7.53}$ & $\ll 1\%$ \\
 & 0.1 -- 0.3 & 0.06 & 152 & 0.59$^{+0.11}_{-0.11}$ & 2.84$^{+0.55}_{-0.55}$ & $\ll 1\%$ \\
 & 0.3 -- 1 & 0.07 & 687 & 0.85$^{+0.08}_{-0.08}$ & 1.48$^{+0.14}_{-0.14}$ & $\ll 1\%$ \\
 & 1 -- 3 & 0.08 & 1094 & 3.34$^{+0.85}_{-0.88}$ & 2.13$^{+0.55}_{-0.55}$ & $\ll 1\%$ \\
 & 3 -- 10 & 0.10 & 569 & 9.42$^{+0.63}_{-0.67}$ & 2.05$^{+0.15}_{-0.15}$ & $\ll 1\%$ \\
 & 10 -- 100 & 0.14 & 134 & 26.35$^{+2.10}_{-2.48}$ & 1.94$^{+0.20}_{-0.20}$ & $\ll 1\%$ \\
\hline
\end{tabular}
\end{table*}

The main motivation of this work is to evaluate whether radio luminosity at low radio frequencies can be used as a SFR indicator, and how our results can be compared with the relations previously obtained using higher radio frequencies.

We fitted models to the all data points of SFGs to determine the relationship between the \magphys\ best fit estimate of SFR and 150-MHz luminosity, including the estimated luminosities of non-detections which were treated in the same way as detections.  The relationship was obtained using  MCMC (implemented in the {\it emcee} {\sc python} package: \citealt{Foreman-Mackey+13}) incorporating the errors on both SFR and $L_{150}$ and an intrinsic dispersion in the manner described by \cite{2010ref7}. Initially we fitted a power law of the form

\begin{equation}
L_{150} = L_1 \psi^\beta, 
\label{eq-1}
\end{equation}

where $\psi$ is the SFR in units of $M_{\odot}$ yr$^{-1}$; $L_1$ has a physical interpretation as the 150-MHz luminosity of a galaxy with a SFR of 1 $M_\odot$ yr$^{-1}$. A Jeffreys prior (uniform in log space) is used for $L_1$ and the form of the intrinsic dispersion is assumed to be lognormal, parametrized by a nuisance parameter $\sigma$. The derived Bayesian estimates of the slope and intercept of the correlation with their errors (one-dimensional credible intervals, i.e. marginalized over all other parameters.) are $\beta = 1.07 \pm 0.01$, $L_1 = 10^{22.06 \pm 0.01}$ W\ Hz$^{-1}$ and $\sigma=1.45 \pm 0.04$.

\begin{figure*}

\begin{center}

\scalebox{1.3}{

\begin{tabular}{c}

\includegraphics[width=11cm,height=11cm,angle=0,keepaspectratio]{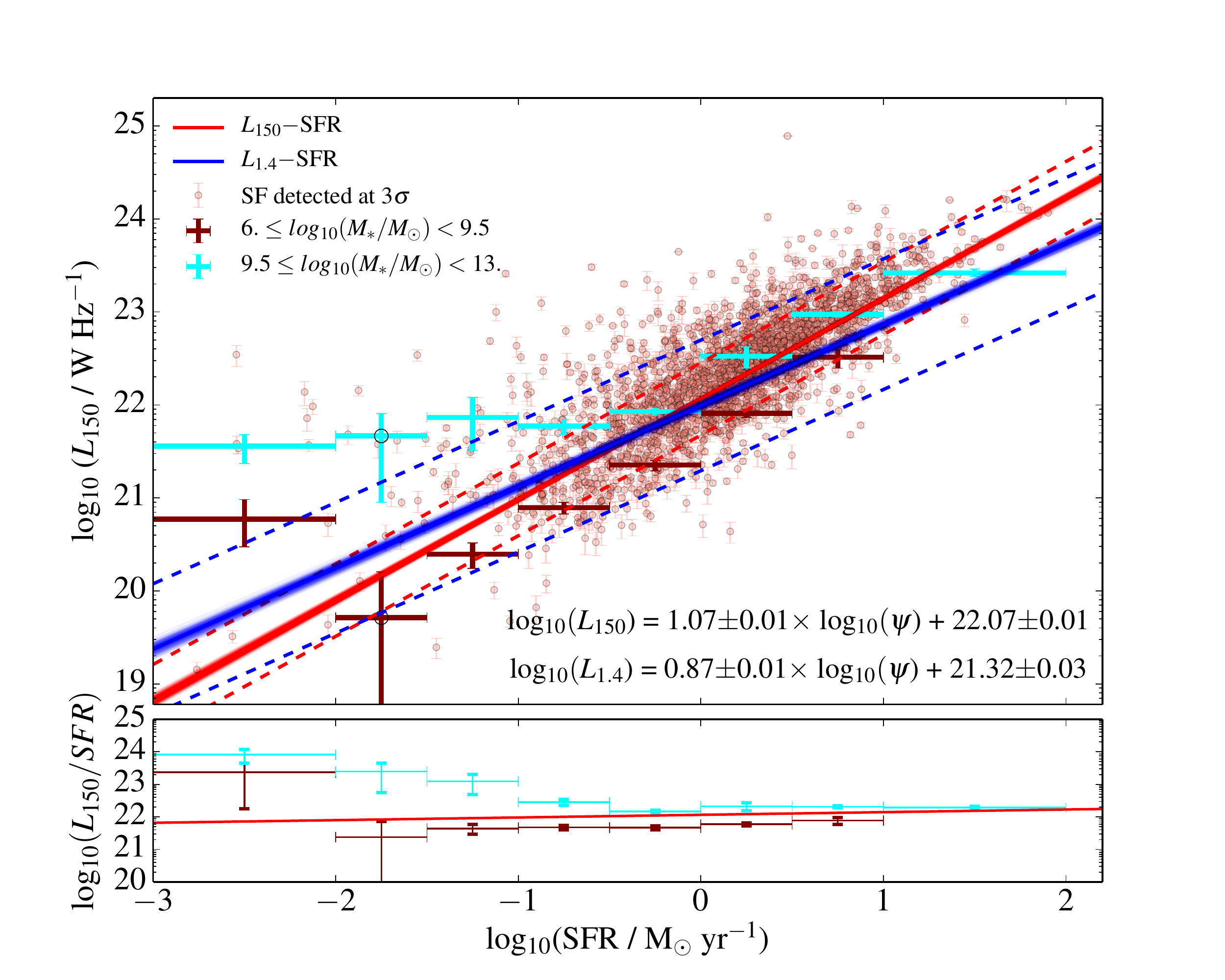}\\
\end{tabular}}
\caption{Top: The distribution of LOFAR 150-MHz luminosities of SFG detected at 150 MHz as a function of their SFRs. The best fit obtained using all SFGs and LOFAR 150-MHz luminosities and errors on the best fit are shown as red shaded region and the blue shaded region shows the best fit to all data points of SFG obtained using $L_{1.4}$ and scaled to 150 MHz assuming $\alpha = 0.8$. The dashed lines around the best fits show the dispersion around the best-fit line implied by the best-fitting dispersion parameter $\sigma$. The results of the stacking analysis for two stellar mass bins are also shown for $L_{150}$ as large cyan and maroon crosses. Open circles indicate the bins in which sources were not detected significantly. Bottom: The $L_{150}$--SFR ratios for two stellar mass bins are shown as cyan and maroon crosses, and the best fit divided by the SFR (red line) are shown.  \label{lfr-sfr}}
\end{center}
\end{figure*}

 To be able to make a consistent comparison with the high-frequency radio luminosity--SFR relation we used the 1.4-GHz luminosities of the same SFGs, derived using FIRST fluxes as described in Section \ref{sec:spix}, and fitted them in the same way (including non-detections). We obtained different values: $\beta_{1.4} = 0.87 \pm 0.01$, $L_1,1.4 = 10^{21.32 \pm 0.03}$ W\ Hz$^{-1}$ and $\sigma=4.02 \pm 0.3$. In the top panel of Fig. \ref{lfr-sfr} shows the distribution of the 150-MHz luminosity of SFGs detected at 150 MHz against their SFRs, together with the best fits that were obtained from the regression analysis.

We investigated the mass-dependence of the $L_{150}$-SFR relation by carrying out a stacking analysis. The $L_{150}$ of SFGs classified by the BPT, independent of whether they were detected at 3$\sigma$ at 150 MHz, were initially divided into two stellar mass bins and then stacked in 8 SFR bins (chosen to have an equal width as well as to have sufficient sources for stacking analysis) using the SFR derived from \magphys. For two stellar mass ranges we determined the weighted average values (treating detections and non-detections together) of the $L_{150}$ samples in individual SFR bins, which are shown as large cyan and maroon crosses in Fig. \ref{lfr-sfr}. Errors are derived using the bootstrap method. The stacking analysis allows us not to be biased against sources that are weak or not formally detected. No fitting analysis was carried out using the stacks but we show them in our figures to allow visualization of the data including non-detections. In the bottom panel of Fig. \ref{lfr-sfr} we show the $L_{150}-$SFR ratios for each SFR bin as large cyan and maroon crosses for two stellar mass bins. The red line shows the best fit, obtained from the regression as described above, divided by SFR.

In order to examine quantitatively whether sources in the bins were significantly detected, we measured 150-MHz flux densities from 100,000 randomly chosen positions in the field and used a Kolmogorov-Smirnov (KS) test to see whether the sources in each bin were consistent with being drawn from a population defined by the random positions. The SFR bins, the number of sources included in each bin and the results of the KS test are given in Table \ref{tbl-lfr-sfr0}. It can be seen from these values that SFG in most bins (except the second SFR bins for both mass ranges: $6.\leq \log_{10}(M_{*})<9.5$ and $9.5\leq \log_{10}(M_{*})<13.$) are significantly detected. Low values of the KS test statistic ($p$-values below 1 per cent) indicate that the target sample in each bin and the randomly selected sources were not drawn from the same distribution, and therefore that the bin is significantly detected. In the bottom panel of Fig. \ref{lfr-sfr} we show the calculated ratios of $L_{150}/\psi$ for the sample of SFGs.

\begin{figure*}

\begin{center}

\scalebox{1.1}{

\begin{tabular}{c}

\includegraphics[width=11cm,height=11cm,angle=0,keepaspectratio]{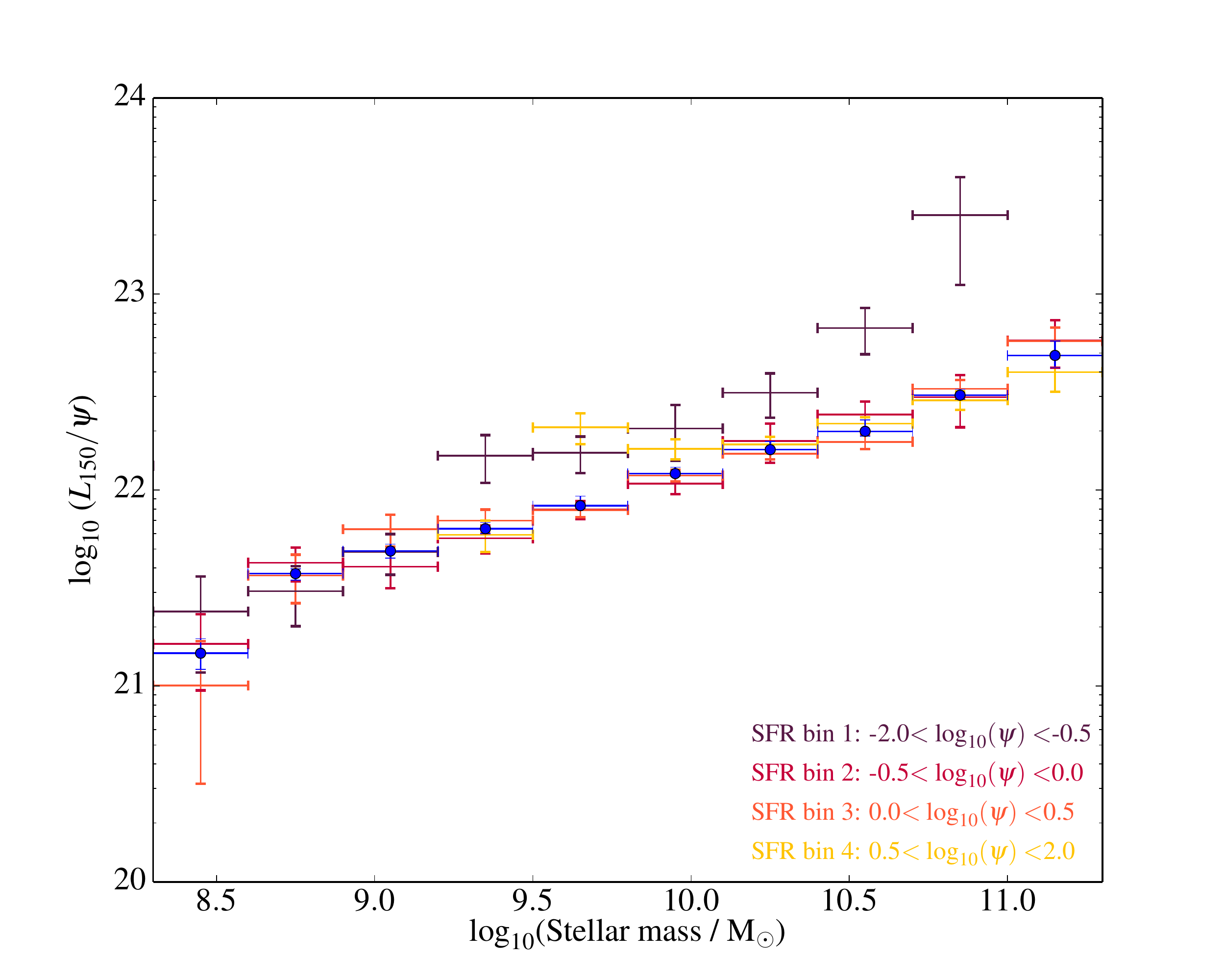}\\
\end{tabular}}
\caption{The results of the stacking analysis of the $L_{150}$--SFR ratios in ten stellar mass bins are shown against stellar mass for the SFG sample. The range of SFR bins and their corresponding colours are shown in the bottom left of the figure. Black filled circles indicate weighted average estimates of stacks in these mass bins. We can see that the $L_{150}$/SFR ratio does not show a dramatic break as a function of stellar mass: instead we see a smooth trend of this ratio with stellar mass.\label{lfr-sfr-mass-bin}}
\end{center}
\end{figure*}

 We next carried out further stacking analysis in order to study the variation of the L$_{150}$/SFR ratio as a function of stellar mass for SFGs in the sample. We initially divided the SFG sample into 4 SFR bins. We then divided each SFG subsample in individual SFR bins into 10 stellar mass bins\footnote{We experimented with several choices of stellar mass binning: our results are robust to the choice of bin boundaries.} and calculated the weighted average of the L$_{150}$/SFR ratio. These ratios, plotted against stellar mass are shown in Fig. \ref{lfr-sfr-mass-bin}. The SFR bin ranges and their corresponding colours are presented in the bottom-right part of the bottom panel.  We also estimated the weighted average of these L$_{150}$/SFR stacks in each stellar mass bin and these are shown as black filled circles. The SFR and stellar mass bins as well as the derived weighted averages of the L$_{150}$/SFR ratio in each are given in Table \ref{ratio_mass_tbl}. We see that all of these analyses suggest a dependence of the radio luminosity on both star formation rate and stellar mass, in the sense that radio luminosity increases with both quantities.

\begin{table}

\caption{Results of stacking $L_{150}$/SFR of SFGs in bins of SFR and stellar mass. Column 1: The range of SFR spanned by the bin. Column 2: The range of stellar mass spanned by the bin. Column 3: The total number of sources in each bin. Column 5: The calculated weighted average of $L_{150}$/SFR. \label{ratio_mass_tbl}}
\begin{tabular}{cccc}
\hline
SFR bins & Stellar mass bins & N & $q\,(L_{150}/SFR)$ \\
$M_{\odot}$ yr$^{-1}$&log$_{10}$($M_{\odot}$)&&($\times
10^{22}$ W Hz$^{-1}$ $M_\odot^{-1}$ yr)\\
\hline
0.03 -- 0.3 & 8.0$-$8.3 & 40 & 1.33$^{+0.84}_{-0.84}$ \\
 & 8.3$-$8.6 & 69 & 0.24$^{+0.12}_{-0.12}$ \\
 & 8.6$-$8.9 & 147 & 0.30$^{+0.10}_{-0.10}$ \\
 & 8.9$-$9.2 & 163 & 0.48$^{+0.11}_{-0.11}$ \\
 & 9.2$-$9.5 & 146 & 1.50$^{+0.41}_{-0.41}$ \\
 & 9.5$-$9.8 & 87 & 1.55$^{+0.32}_{-0.32}$ \\
 & 9.8$-$10.1 & 56 & 2.06$^{+0.65}_{-0.65}$ \\
 & 10.1$-$10.4 & 31 & 3.13$^{+0.81}_{-0.81}$ \\
 & 10.4$-$10.7 & 14 & 6.71$^{+1.77}_{-1.77}$ \\
 & 10.7$-$11.0 & 10 & 25.26$^{+13.48}_{-13.48}$ \\
0.3 -- 1.0 & 8.0$-$8.3 & 20 & -0.01$^{+0.05}_{-0.05}$ \\
 & 8.3$-$8.6 & 29 & 0.16$^{+0.07}_{-0.07}$ \\
 & 8.6$-$8.9 & 32 & 0.43$^{+0.08}_{-0.08}$ \\
 & 8.9$-$9.2 & 128 & 0.41$^{+0.09}_{-0.09}$ \\
 & 9.2$-$9.5 & 178 & 0.57$^{+0.10}_{-0.10}$ \\
 & 9.5$-$9.8 & 233 & 0.79$^{+0.09}_{-0.09}$ \\
 & 9.8$-$10.1 & 221 & 1.08$^{+0.12}_{-0.12}$ \\
 & 10.1$-$10.4 & 142 & 1.78$^{+0.41}_{-0.41}$ \\
 & 10.4$-$10.7 & 59 & 2.43$^{+0.44}_{-0.44}$ \\
 & 10.7$-$11.0 & 25 & 2.97$^{+0.84}_{-0.84}$ \\
 & 11.0$-$11.3 & 7 & 5.79$^{+1.67}_{-1.67}$ \\
1.0 -- 3.0 & 8.3$-$8.6 & 8 & 0.10$^{+0.06}_{-0.06}$ \\
 & 8.6$-$8.9 & 17 & 0.37$^{+0.11}_{-0.11}$ \\
 & 8.9$-$9.2 & 37 & 0.63$^{+0.12}_{-0.12}$ \\
 & 9.2$-$9.5 & 105 & 0.70$^{+0.10}_{-0.10}$ \\
 & 9.5$-$9.8 & 212 & 0.79$^{+0.07}_{-0.07}$ \\
 & 9.8$-$10.1 & 290 & 1.19$^{+0.08}_{-0.08}$ \\
 & 10.1$-$10.4 & 299 & 1.53$^{+0.10}_{-0.10}$ \\
 & 10.4$-$10.7 & 208 & 1.76$^{+0.13}_{-0.13}$ \\
 & 10.7$-$11.0 & 68 & 3.28$^{+0.36}_{-0.36}$ \\
 & 11.0$-$11.3 & 15 & 5.75$^{+1.03}_{-1.03}$ \\
3.0 -- 32. & 9.2$-$9.5 & 22 & 0.59$^{+0.11}_{-0.11}$ \\
 & 9.5$-$9.8 & 64 & 2.09$^{+0.38}_{-0.38}$ \\
 & 9.8$-$10.1 & 99 & 1.62$^{+0.20}_{-0.20}$ \\
 & 10.1$-$10.4 & 154 & 1.71$^{+0.16}_{-0.16}$ \\
 & 10.4$-$10.7 & 202 & 2.18$^{+0.17}_{-0.17}$ \\
 & 10.7$-$11.0 & 138 & 2.87$^{+0.30}_{-0.30}$ \\
 & 11.0$-$11.3 & 41 & 3.99$^{+0.85}_{-0.85}$ \\
\hline
\end{tabular}
\end{table}

 Galaxy mass has an important role, in particular for non-calorimeter models, in the relation of $L_{\mathrm{radio}}$ and $\psi$ because the galaxy size has an impact on the competition between radiative loss and \cre\ diffusion \citep[e.g.][]{2003lf10,2010lf19,2010lf20}. In order to quantify the role that stellar mass plays we fitted the data taking into account both quantities, using the following empirical parametrization:

\begin{equation}
L_{150} = L_C \psi^{\beta}\left(\frac{M_{*}}{10^{10}M_{\odot}}\right)^{\gamma}
\label{eq-2}
\end{equation}

 We find $\beta=0.77\pm$0.01, $\gamma=0.43\pm$0.01, normalization $L_C=10^{22.13(\pm0.01)}$ W Hz$^{-1}$, where $L_C$ is now the luminosity of a galaxy with $M_{*} = 10^{10}{\mathrm M}_{\odot}$ and $\psi = 1\ {\mathrm M}_\odot$ yr$^{-1}$ and the intrinsic scatter $\sigma=1.71\pm0.05$. We utilized Petrosian radius (petroR50) measurements provided by SDSS to derive galaxy sizes $R_{e}$ (effective half-light radius) for our SFGs as these were used in order to obtain the relation between $L_{150}$, SFR and galaxy size using the same parameterization given in Equation \ref{eq-2} (see Table \ref{tbl-lfr-sfr}).

 Quantitative evaluation of the two parameterizations used in the regression analyses is important in order to understand which parameterization is favoured by the data. For this we carried out Bayesian model selection. A comparison of the Bayesian evidence (i.e.\ the integral of the likelihood over parameter space) shows that the model with mass dependence is strongly favoured over the one without (as also indicated in Figs. \ref{lfr-sfr} and \ref{lfr-sfr-mass-bin}). Very similar results, which for brevity we do not discuss here, are found using a measure of galaxy physical size instead of mass, which is not surprising since the two are very tightly correlated \citep[e.g.][who show that $R_{e}$ $\propto M_{*}^{0.1-0.4}$]{2015lange}. The best-fitting model parameters for this model are given in Table \ref{tbl-lfr-sfr}.

 Visual inspection of Fig.\ \ref{lfr-sfr}, and in particular the results of the stacking analysis, indicated that sources with low SFRs might be parametrized by different power laws, compared with sources with high SFRs, as previously proposed by \cite{1990chi} and \cite{2003lf10}. In addition to this, studies of the SFR function and the local radio luminosity function also indicate a difference between SFR of low stellar mass galaxies and high stellar mass galaxies \citep[e.g.][]{Mancuso15,bonato17,massardi10}. In order to investigate this further we fitted the data with a broken power law, of the form:

\begin{equation}
L_{150} = \left\{ \begin{array}{ll} L_C
    \psi^{\beta_{low}}\left(\frac{M_{*}}{10^{10}M_{\odot}}\right)^{\gamma_{2}}&\psi
    \le \psi_{\rm break}\\
    L_C
    \psi^{\beta_{high}}\left(\frac{M_{*}}{10^{10}M_{\odot}}\right)^{\gamma_{2}}
    \psi_{\rm break}^{(\beta_{low}
        - \beta_{high})}&\psi
    > \psi_{\rm break}\\
    \end{array}\right .
\label{broken}
\end{equation}
where $\psi_{\rm break}$ is the SFR at the position of the break, a free parameter of the fit.

 A plot of the results of this analysis is shown in Fig. \ref{broken}; here the mass-dependent effect has been taken out so that the star-formation -- radio- luminosity relation alone can be seen. The results of the regression analysis suggest that there is a break in SFR around 1.02 M$_{\odot}$ yr$^{-1}$. SFGs with SFRs higher/lower than this value favour different parameterizations (see Table \ref{tbl-lfr-sfr} for derived parameters). Again, the broken power-law model is strongly favoured on a Bayesian model comparison over the single power law with a mass dependence. We implemented the same regression analysis using high frequency radio measurements at 1.4 GHz (the best fit obtained using all SFGs, and the uncertainties, by overplotting the lines corresponding to a large number of samples from the MCMC output are shown with blue colour in Fig. \ref{broken}). The physical interpretation of these results and their comparison with the literature are discussed in Section \ref{sec:results}.

As can be seen in the upper panel of Fig. \ref{lfr-sfr} and in Fig. \ref{broken} there are LOFAR-detected sources with low SFRs ($-3<\log_{10}(\psi)<-1$) and high $L_{150}$, that lie off the `main sequence'. We visually inspected these sources in order to make sure that these objects were not blended radio sources or sources with uncertain redshift estimates. The inspections showed that these are genuine SFGs at $0.02<z<0.13$. We excluded these objects from the SFG sample and implemented the same MCMC regression analysis using the broken power-law parameterization. The results showed that the fit is unaffected by excluding these objects.

\begin{table*}

\caption{Results of various parameterizations and derived parameters from our regression analyses. Given errors are one-dimensional credible intervals, i.e. marginalized over all other parameters.
 \label{tbl-lfr-sfr}}
	\begin{tabular}{lrrrrrrr}
		\hline
Parametrization & $\beta$ & $\gamma$  & $\beta_{low}$ & $\beta_{high}$ & $\log_{10}(\psi_{break})$ & Normalization&$\sigma$\\
		\hline
$L_{150} = L_C \psi^{\beta}$& 1.07$\pm$0.01 & -& -& -& -& 22.06$\pm$0.01& 1.45$\pm$0.04\\
$L_{150} = L_C \psi^{\beta}\left(\frac{M_{*}}{10^{10}M_{\odot}}\right)^{\gamma}$& 0.77$\pm$0.01& 0.43$\pm$0.01& -& -& -& 22.13$\pm$0.01& 1.71$\pm 0.05$\\
$L_{150} = L_C \psi^{\beta}\left(\textbf{R}_{e} [pc]\right)^{\gamma}$& 0.95$\pm$0.01& 0.51$\pm$0.02& -& -& -& 22.16$\pm$0.01& 1.37$\pm$0.04\\
Broken power-law (using $L_{150}$ $\psi \le \psi_{\rm break}$) & -& 0.44$\pm 0.01$& 0.52$_{-0.03}^{+0.03}$& -&-& 22.02$\pm$0.02 & $1.68 \pm 0.05$\\
Broken power-law (using $L_{150}$ $\psi > \psi_{\rm break}$)& -& 0.44$\pm 0.01$& 0.52$_{-0.03}^{+0.03}$& 1.01$ \pm 0.02$ & 0.01$ \pm 0.01$& 22.02$\pm$0.02& $1.68 \pm 0.05$\\
Broken power-law (using $L_{1.4}$ $\psi \le \psi_{\rm break}$) & -& 0.40$\pm 0.01$& 0.48$\pm 0.03$& -&-& 21.35$\pm$0.03 & 4.12$_{-0.36}^{+0.25}$\\
Broken power-law (using $L_{1.4}$ $\psi > \psi_{\rm break}$)& -& 0.40$\pm 0.01$& 0.48$\pm 0.03$& 0.85$_{-0.03}^{+0.02}$& 0.54$_{-0.08}^{+0.07}$&21.35$\pm$0.03 & 4.12$_{-0.36}^{+0.25}$\\
\hline
\end{tabular}
\end{table*}

\begin{figure*}
\scalebox{1.5}{
\begin{tabular}{c}
\hspace{-2em}\includegraphics[width=11cm,height=11cm,angle=0,keepaspectratio]{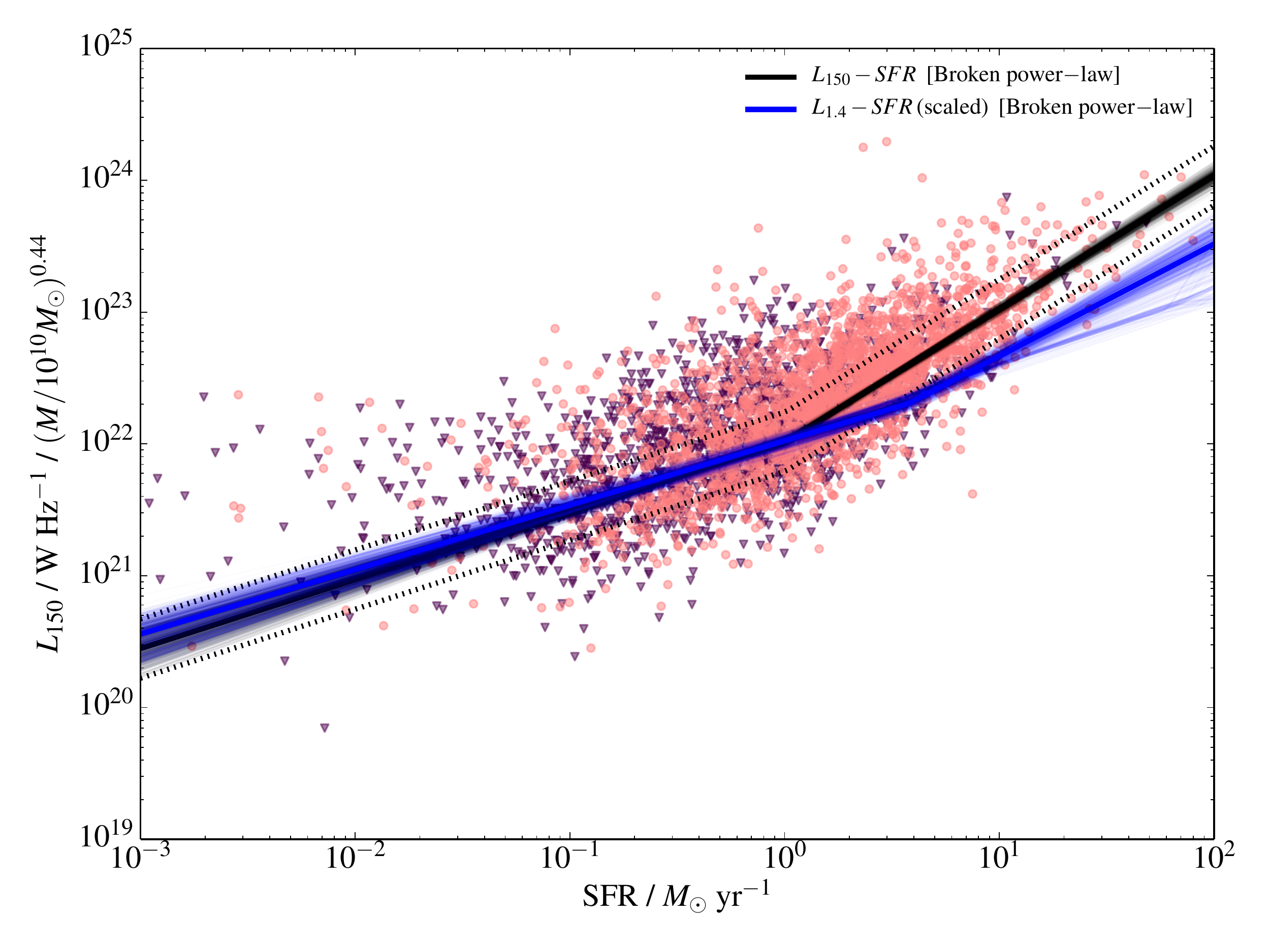}\\
\end{tabular}}
\caption{Distribution of $L_{150}$ of SFGs as a function of their SFRs. The mass-dependent effect is taken out (by dividing $L_{150}$ by the fitted mass-dependent term) so that the star-formation -- radio-luminosity relationship can be seen. Upper limits are indicated with purple triangles. The best fit obtained using all SFGs, and the uncertainties, are visualized by overplotting the lines corresponding to a large number of samples from the MCMC output. The dashed black lines show the $1\sigma$ intrinsic dispersion around the best-fit line implied by the best-fitting dispersion parameter $\sigma$. The same relation derived using high frequency observations is shown as the blue line, with fitting uncertainties displayed in the same way. \label{broken}}
\end{figure*}


\subsection{The far-IR--radio correlation}

\begin{figure*}
\scalebox{1.3}{
\begin{tabular}{c}
\includegraphics[width=11cm,height=11cm,angle=0,keepaspectratio]{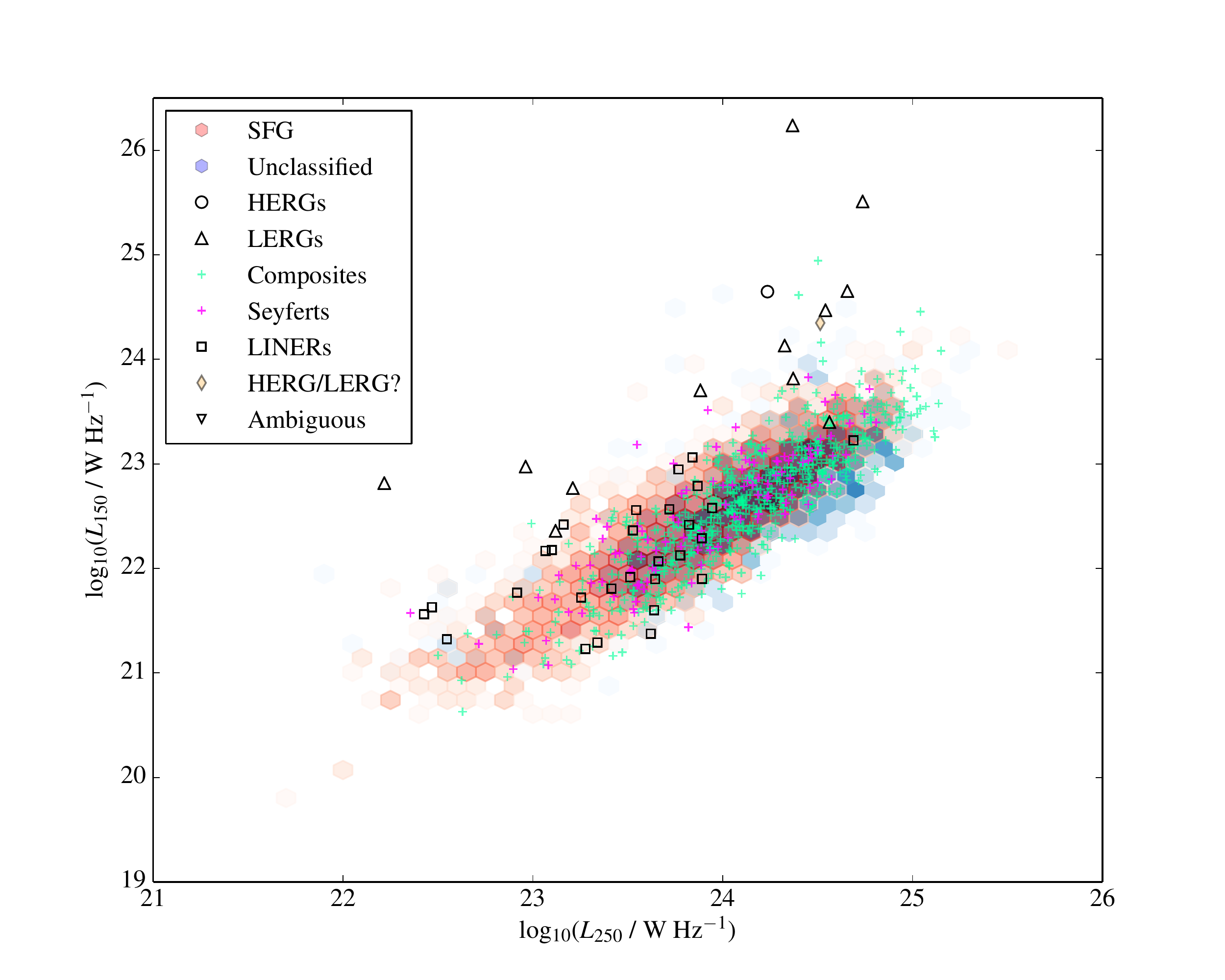}\\
\end{tabular}}
\caption{The distribution of LOFAR 150-MHz luminosities of different classes of objects in the sample detected at 3$\sigma$ as a function of their $\textit{Herschel}$ 250-$\mu$m luminosities. Symbols and colours as for Fig. \ref{lum}.}
\label{fir-rad-all}
\end{figure*}

The data also allow us to investigate the low frequency radio -- far-IR correlation, originally presented for LOFAR-detected sources in this sample by H16. In Fig.\ \ref{fir-rad-all} the radio-luminosity--far-IR luminosity correlation for all sources detected by both {\it Herschel} and LOFAR is plotted. Unlike the corresponding SFR figure (Fig.\ \ref{lfr-sfr-all}) all the sources detected in both bands lie on what appears to be a good correlation: this is partly a selection effect in that most of the unclassified sources detected by LOFAR are not detected by {\it Herschel} and so do not appear on the figure. In what follows we investigate this correlation using only sources classed as SFGs (4157 sources) using the emission line classification. We note that for these sources the correlation between $L_{150}$ and $L_{250}$ appears tighter than that between $L_{150}$ and SFR, which, assuming that the scatter in the latter is not dominated by unknown errors in the \magphys-derived SFR, illustrates the well-known `conspiracy' between FIR and radio luminosity.

In order to find the relation between $L_{150}$ and $L_{250}$ we fitted a power-law model using MCMC in the same way, as explained in the previous section, fitting a power-law model of the form $L_{150} = C L_{250}^\beta$. We used all SFGs in our fitting process, whether detected by LOFAR or {\it Herschel} or not -- this procedure should give us an unbiased view of the true correlation.

\begin{figure*}
\begin{center}
\scalebox{1.5}{
\begin{tabular}{c}
\hspace{1em}\includegraphics[width=11cm,height=11cm,angle=0,keepaspectratio]{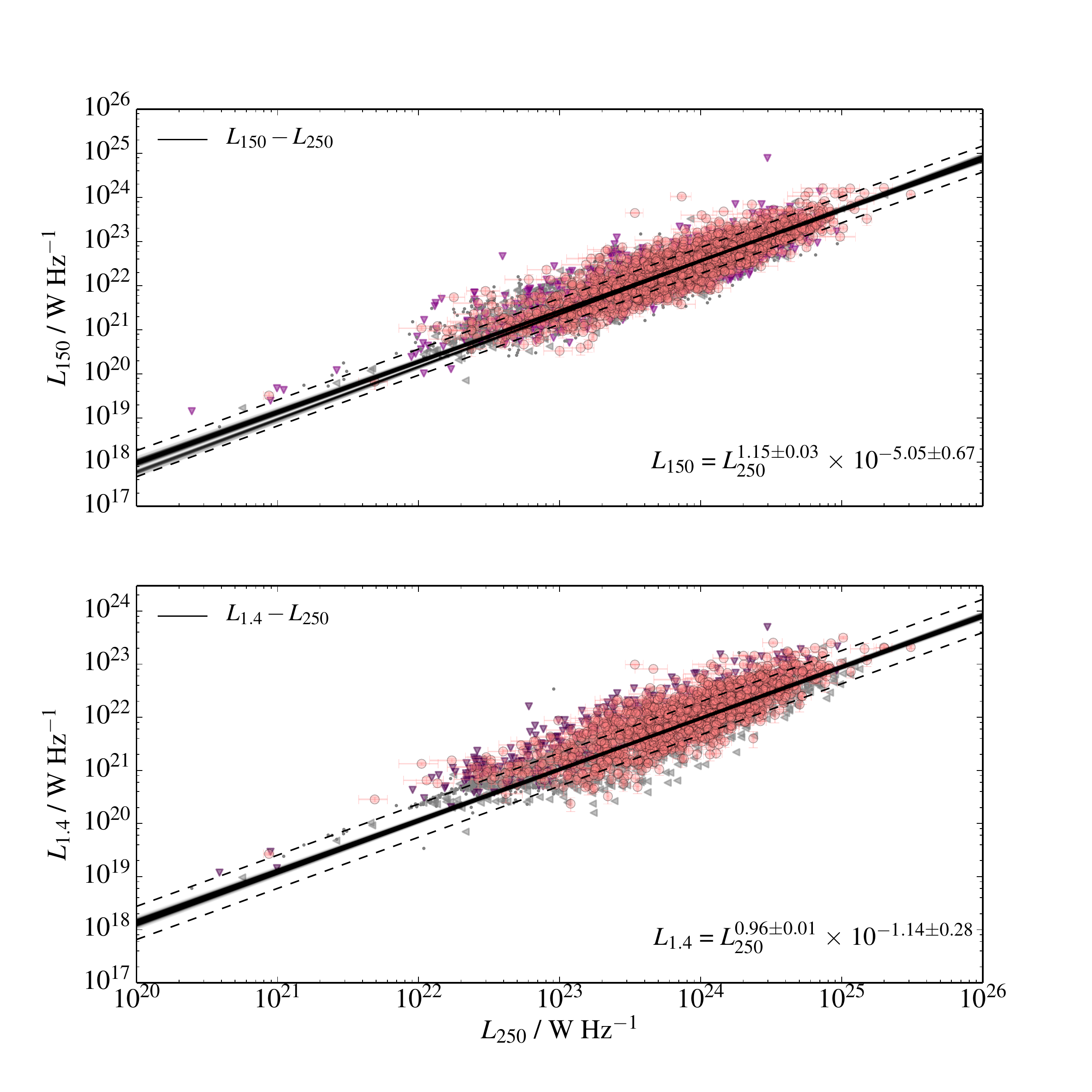}\\
\end{tabular}}
\caption{Top: The distribution of ${\it Herschel}$ 250-$\mu$m luminosities of SFGs as a function of their LOFAR 150-MHz luminosities. As for Fig.\ \ref{broken}, the best-fit line and its dispersion are visualized by plotting the lines corresponding to many MCMC samples. Dashed lines indicate the $1\sigma$ intrinsic dispersion of the fit parametrized by the nuisance parameter $\sigma$. Salmon circles show 3$\sigma$ detections; LOFAR 150-MHz $3\sigma$ limits are indicated by purple down-pointing triangles and $3\sigma$ limits on ${\it Herschel}$ 250-$\mu$m luminosities are indicated by grey left-pointing triangles; grey dots indicate $3\sigma$ upper limits on both quantities. A tight relationship between $L_{250}$ and $L_{150}$ is clearly seen for local SFGs. Bottom: The distribution of ${\it Herschel}$ 250-$\mu$m luminosities of SFGs as a function of their FIRST 1.4-GHz luminosities. Symbols and lines as for the top panel. \label{fir-rad}}
\end{center}
\end{figure*}

The derived Bayesian estimate of the slope and intercept of the correlation are:

\begin{equation}
L_{150}=L_{250}^{1.15(\pm0.02)}\times 10^{-5.05^{+0.34}_{-0.77}}
\end{equation}

with the best-fitting dispersion parameter $\sigma=0.3\pm0.02$. The results of the regression analysis are shown in Fig.\ \ref{fir-rad} where we plot the distribution of $L_{250}$ of all SFGs against their $L_{150}$ together with $\pm 1\sigma$ errors on both best fits.

The same fitting procedure was carried out using $L_{1.4}$ to obtain the $L_{1.4}-L_{250}$ relation. This allowed us to make a consistent comparison of the relations derived using radio luminosities at different radio frequencies for the same sample. Fitting was implemented in the manner discussed above using MCMC; the best-fitting relations derived from this are as follows:

\begin{equation}
L_{1.4}=L_{250}^{0.96(\pm 0.01)}\times 10^{-1.14 \pm 0.27}
\end{equation}

with $\sigma=1.05 \pm 0.03$. We see that $L_{1.4}$ gives a slightly shallower slope than $L_{150}$. This is consistent with the fact that a shallower slope was also observed in the $L_{1.4}$--SFR relation derived in the previous section.


\section{Interpretation}

\label{sec:results}

\subsection{The $L_{150}$--SFR relation}

\label{sec:l150-sfr}

\subsubsection{Slope and variation with SFR}

The slope obtained from the regression analysis of $L_{150}$ against SFR using SFGs is close to, but slightly steeper than unity ($\beta=1.10\pm0.01$). In the bottom panel of Fig. \ref{lfr-sfr} we show the ratio of the mean of $L_{150}$ to SFR in each stellar mass and SFR bin (see Table \ref{tbl-lfr-sfr} for the numerical values of these ratios) as well as the regression line (the solid red line in the top panel) divided by SFR. These stacks show that the relation between SFR and non-thermal emission from SFGs as a function of SFR is almost constant for sources with high radio luminosities (or around $\psi>1$) whereas low-luminosity sources present slightly higher ratios. More importantly, the results of the stacking analysis in the top panel of Fig. \ref{lfr-sfr} show that for two mass bins the relation is different for low-SFR SFGs and high-SFR galaxies. The results of the stacking analysis and our broken power-law fits both indicate that low-SFR and high-SFR SFGs diverge from one another in terms of the $L_{150}$--SFR relation. Furthermore, in Fig. \ref{lfr-sfr-mass-bin} we can see that the $L_{150}$/SFR ratio does not show dramatic break as a function of stellar mass: instead we see a smooth trend of this ratio with stellar mass.

Our sample consists of SFGs selected in a consistent way (using optical emission lines) and covers a narrow redshift range, so we do not expect to see strong cosmological effects. So why do SFGs with different luminosities (or SFR) differ from each other in terms of the $L_{150}-SFR$ relation? An interesting feature of this difference is that the 150-MHz luminosity of low-SFR galaxies is generally {\it higher} than would be predicted from a fit to the high-SFR galaxies alone (see Fig.\ \ref{broken}).  The difference is therefore in the opposite sense to the predictions of models which postulate that the electron calorimetry approximation breaks down for low SFR and would therefore predict a {\it deficiency} in radio luminosity for low SFR \citep[e.g.][Kitchener et al. in press]{1991lf45,1992lf41,1996lf34,2003lf10}.  The mass dependence of luminosity is in the sense that more massive galaxies are more luminous, and this {\it is} in the sense predicted by such models, since more massive galaxies are larger and so have longer escape times. However, the flatter slope of the $L_{150}$ -- SFR relation at low SFR is not. One possibility is that there is another source of the cosmic rays radiating at in these low-SFR, low-mass galaxies, such as pulsars or type Ia supernovae, and that what we are seeing going from high to low star formation rates is a transition from a regime in which the radio luminosity depends almost exclusively on the SFR in a calorimeter model, through one where the calorimeter assumptions break down but star formation is still driving cosmic ray generation, down to one where other sources of cosmic rays take over and star formation is irrelevant. Another possibility would be to invoke amplification of magnetic fields in these high-SFR galaxies (probably) due to different dynamos or galactic winds. Testing such scenarios in detail would require a much better model of the relevant galaxy-scale physics, as well as more data (more sensitive data over larger fields).

\subsubsection{Frequency dependence of the relation}

\label{sec:fir-discus}

The slope of the $L_{1.4}-\psi$ relation ($\beta=0.87$) derived using $L_{1.4}$ is lower than the slope ($\beta=1.07$) obtained using $L_{150}$. Steepening of the slope towards to low frequencies has been previously observed in studies of the FIRC \cite[e.g.][]{1992lf41,1997lf42}, and these authors argued that thermal radio emission comes to dominate at higher radio frequencies whereas non-thermal radio emission dominates at lower radio frequencies, giving a steeper correlation with SFR. This is because thermal radio emission has a flat spectrum, $S_\nu \propto \nu^{-0.1}$ whereas the spectrum is steeper for non-thermal radio emission ($S_\nu \propto \nu^{-0.7}$). Radio luminosity from SFGs at 150 MHz should have a negligible level of contamination from thermal radio emission.  However, at 1.4 GHz we might expect to observe some contribution from thermal emission \citep{1992lf5}. Therefore, the different slopes might be attributed to differing contributions from thermal emission at 150 MHz and 1.4 GHz. A much more detailed radio SED for each galaxy would be needed to test this model. The same picture is also seen when we used a broken power-law parameterization: both slopes obtained using the $L_{1.4}$ are shallower than the slopes estimated from the $L_{150}$ luminosities. On the other hand, the best fitting line that we obtain for low SFR SFGs indicates higher luminosities than what we actually derive as a slope using 150 MHz luminosities of these sources. This might be due to the effects of synchrotron self absorption. We are not able to test this with our currently available data. However, we plan to investigate this further using follow up observations. Another interesting point is that the SFR-break values that we obtain using low and high frequency observations are different (see Table \ref{tbl-lfr-sfr}). This might be again due to a different balance between the thermal and non-thermal emission that we observe at low and high frequencies.

\subsubsection{Comparison to the literature}

 Although the majority of authors still investigate it in terms of the FIRC, the explicit relation between radio luminosity and SFR in both normal star-forming galaxies and starbursts has been studied a number of times previously \citep[e.g.][]{1992lf5,Cram+98,2001lf9,2003lf10,Hodge+08,2009lf43,murphy+11,taba+17}. Key points of difference in our analysis are (i) we start from a sample of galaxies selected using optical emission lines and include radio and FIR non-detections in all regressions, rather than selecting in the radio; (ii) we work at the unusually low frequency of 150 MHz, thus reducing the effects of thermal radio emission; (iii) we probe down to low SFRs with our local sample; (iv) we use energy-balance SED fitting to derive galaxy properties; and (v) we have tried to incorporate other information about the galaxies by including a mass- dependent term in our regression.

 Considering first the calibration of the 150-MHz-SFR relation, we find that the overall power-law normalization of our result agrees well with determinations by \cite{Calistro17}. As mentioned previously, they use a radio detected sample and investigate the relation $L_{150}-SFR$ for sources that cover a wide redshift range ($z<2.5$).  They report a slope of 1.54 which is close to our results but it is steeper than $\beta=1.07$ which we find in our study. It is worth noting that the difference is that we have a much larger sample of galaxies, allowing us to investigate more complex models for the radio--SFR relation.

 Some authors \citep[e.g.][]{2001lf9,2009lf43} have {\it assumed} a linear correlation between SFR and radio emission in calibrating the radio--SFR relation, but \cite{Hodge+08}, who explicitly fitted the correlation, found a super-linear relation, $L \propto \psi^{1.37\pm0.02}$ for SFGs at similar redshifts. If we assume that SFGs lie on the main sequence with $\psi \propto M$, this appears broadly consistent with our fits to the high-SFR objects ($L \propto \psi^{1.01} M^{0.44}$, see table \ref{tbl-lfr-sfr} for SFGs with SFRs$>\psi_{break}$) and is also consistent with recent predictions based on the assumption that the magnetic field strength scales with SFR \citep{Schleicher+Beck13}. Recently, \cite{davies+17} investigated the relation between SFR and 1.4 GHz luminosities for a sample of sources detected in the FIRST survey. They also find a slope less than unity but it ($\beta=0.75$). \cite{brown17} present a study of calibrated SFR indicators using radio continuum emission at both 1.4 GHz and 150 MHz of local star-forming galaxies spanning a similar SFR range to ours, and also find power-law indices steeper than unity for their $L$/$\psi$ relation ($\beta=1.14 \pm 0.05$) which is consistent with our findings ($\beta=1.07 \pm 0.01$).

 \cite{2003lf10}, in their study of the radio--FIR relation in local star- forming galaxies, inferred a break in the radio--SFR relationship for low-SFR (low-radio luminosity) galaxies, in the sense that they should have lower radio emission for a given SFR. It is important to note that this break was not directly observed but rather inferred by \cite{2003lf10} based on the lack of variation in the radio--FIR ratio parameter $q$; as we discuss in the introduction, the interpretation of this parameter is seriously affected by selection effects.  Such a deficiency might well be expected due to the escape of \cre\ in non-calorimeter models \citep{2010lf19}. However, as noted above, we do not observe a radio luminosity deficiency due to the escape of \cre\ in these objects. The break we see for low-SFR objects ($< 1.02 {\mathrm M}_{\odot}$ yr$^{-1}$) is in the sense that these have {\it more} radio emission than would be predicted from an extrapolation of the super-linear trend from higher SFR. This has not been observed before, and is also not seen in the stacking analysis of \cite{Hodge+08}, which may indicate that it is more important to taking into account the mass effect in the relation. Further investigation of these low-SFR SFGs is required.

 Finally, it is important to note that comparison with other work is made more difficult by the fact that study design effects (sample selection, sample size, range of SFRs and redshifts covered, SFR indicators used etc.) play a crucial role in the resulting $L_{\mathrm{radio}}$--SFR relation. In particular we note that, as \cite{Calistro17} show in their LOFAR-based study of a sample of SFGs selected from the Bo\"otes field, the $L_{150}$--SFR relation is likely to vary with redshift, so fits that span a large redshift range (like those of \citealt{2009lf43}) may not recover the intrinsic relationship.

\subsection{The far-IR/radio correlation of SFGs}

 We have derived the radio-FIRrelation at 150MHz for a large sample for the first time in this analysis. At 150 MHz, a note worthy feature (Fig.\ref{fir-rad}) is that the slope of the $L_{150}-L_{250}$ relation is lower than unity ($L_{250}\propto L_{150}^{0.73\pm0.01}$), while the slope is consistent with unity ($0.96\pm 0.02$) for the $L_{1.4}$ of the same galaxies. In terms of the slope of the 150-MHz relationship, our results are similar to those of \citet{1988lf52}, who found $L_{FIR}\propto L_{151}^{0.87\pm0.03}$,  while results of other studies of the correlation at 1.4GHz indicate slopes of unity for this relation for SFGs\citep[e.g.][]{Jarvis+10,Smith+14}.

We interpret the difference between the two relations for our sample, as for the $L_{150}$--SFR relation, in terms of the impact of thermal emission from SFGs observed at 1.4 GHz. Since the low-frequency observations are probably uncontaminated by thermal emission, the roughly constant ratio between radio and far-IR at 1.4 GHz must then be the result of another `conspiracy' which happens to operate in the local universe at that particular combination of radio and IR wavelengths.

The ratio between the two luminosities is often parametrized \citep{Helou+85,Smith+14} in terms of a ratio of monochromatic luminosities, $q = \log_{10}(L_{\rm IR}/L_{\rm radio})$ (or its equivalent in terms of flux density or broad-band IR flux). Here we use $L_{250}$ as our IR luminosity. Considering first $q_{1.4}$ (i.e. $\log_{10}(L_{250}/L_{1.4})$, and computing $q$ at a reference value of $L_{250} = 10^{24}$ W Hz$^{-1}$, we see that the value from the regression line is $\sim 1.95 \pm 0.2$, comparable to the values $\sim 2.0$ reported by e.g. \cite{Jarvis+10}, \cite{Ivison+10} based on observations of sources detected in both radio and IR. \cite{Smith+14} report a larger value, $\sim 2.5$, based on FIRST and H-ATLAS data, but this is most likely because we fit to {\it all} star-forming galaxies, not just to those detected in the far-IR, while Smith et al considered only IR-detected sources. The significant intrinsic dispersion seen in the data (Fig.\ \ref{fir-rad}) means that selection on IR-detected sources will naturally bias values of $q$ high.

If we interpret the 150-MHz best-fit relationship in terms of the parameter $q_{150} = \log(L_{250}/L_{150})$, then we find a more interesting result; $q$ is no longer even approximately constant with $L_{250}$, in the sense that low-luminosity IR sources have much lower ratios of radio to IR luminosity. $q_{150}$ is $1.5 \pm 0.2$ at $L_{250} = 10^{24}$ W Hz$^{-1}$ but $1.7 \pm 0.2$ at $L_{250} = 10^{23}$ W Hz$^{-1}$. It is probably this non-linear dependence of low-frequency luminosity on IR luminosity that accounts for the discrepancy between the number of LOFAR and {\it Herschel} sources discussed by H16.

\subsection{Nature of the objects unclassified by emission lines}

\label{sec:unclassified}

\begin{figure*}
\begin{center}
\scalebox{1.3}{
\begin{tabular}{c}
\includegraphics[width=11cm,height=11cm,angle=0,keepaspectratio]{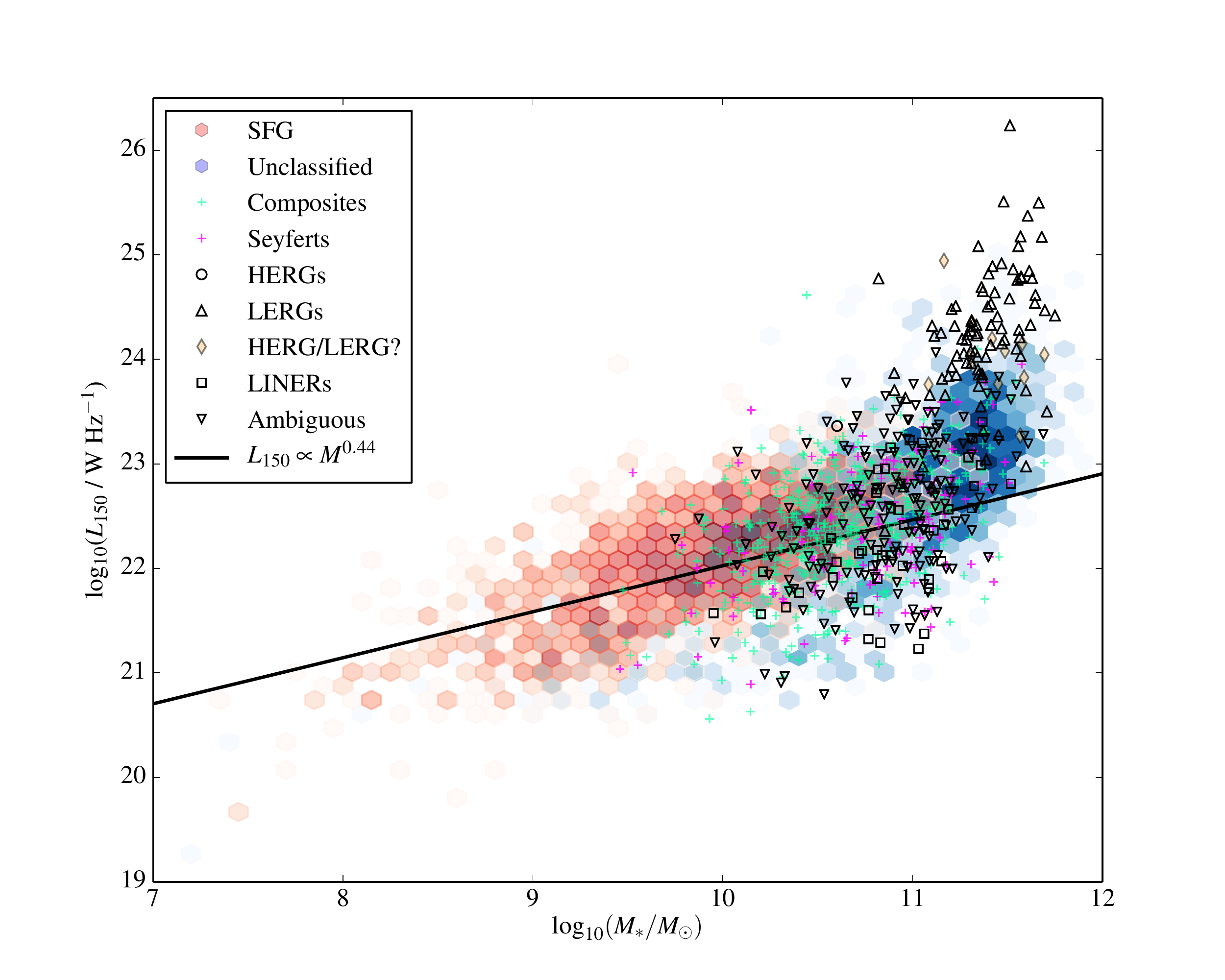}\\
\end{tabular}}
\caption{Radio luminosity for low-SFR LOFAR-detected objects as a function of mass. Symbols and colours as for Fig. \ref{lum}, but to simplify the figure only sources with best-fitting $\log_{10}(\psi)<0.01$ are plotted. The solid line shows the relation between $L_{150}$ and mass, derived from the fits for low-SFR SFGs, for objects with $\log_{10}(\psi) = 0.01$. Many of unclassified sources with higher radio luminosities appear much less anomalous in this figure: they follow an extrapolation of the $L_{150}$--stellar mass relation observed for SFGs, and we see them lying in the same region of the figure as SFGs, LINERs and Composites.}
\label{l150-mass}
\end{center}
\end{figure*}

 As noted above, we saw in Fig. \ref{lfr-sfr-all} that sources that are not strongly (at $3\sigma$) detected in their optical emission-lines do not follow the same relation between $L_{150}$ and SFR as SFGs. Inference of the SFR from the radio luminosity alone would greatly overestimate its value in these objects (assuming our \magphys-derived values are correct). Very similar conclusions were reached, using stacking of FIRST luminosities for the whole MPA-JHU sample, by \cite{Hodge+08}.

 In the light of the discussion in Section \ref{sec:l150-sfr} it is interesting to plot the radio luminosity of these objects against stellar mass: see Fig.\ \ref{l150-mass}. Many of these sources appear much less anomalous in this figure: they follow an extrapolation of the $L_{150}$--stellar mass relation observed for SFGs, and we see them lying in the same region of the figure as SFGs, LINERs and Composites. However, above a stellar mass of $\sim 10^{11}\ M_\odot$ we see an abrupt increase in the typical radio luminosity for a given stellar mass, which culminates in the BH12 radio-loud AGN, which are both the most radio-luminous and the most massive objects in the sample. This is a very clear indication that there is an additional contaminating population of low-luminosity radio-loud AGN, not identified as such by BH12 because of the radio flux density cut they applied (but expected from the shape of their luminosity function). The slightly flatter radio spectrum of the unclassified objects noted in Section \ref{sec:spix} would be consistent with the presence of a compact AGN in some fraction of the population. The combination of the strong mass dependence of (what we take to be) non-AGN radio emission from normal galaxies {\it and} the presence of radio-loud AGN activity at significant levels in essentially all massive galaxies means that a simple inference of star-formation rates from radio luminosity alone is extremely complicated.

\section{Conclusions and future work}

We have analyzed the radio emission from galaxies in the MPA-JHU sample using a LOFAR 150-MHz survey of the H-ATLAS/NGP field (H16). This has allowed us to explore the low-frequency radio luminosity--SFR relation of SFGs in the local Universe ($0<z<0.35$). The SFRs of these objects were derived using SED fitting with \magphys, using SDSS, {\it WISE} and {\it Herschel} data. The \textit{Herschel} 250-$\mu$m flux densities allowed us to investigate the FIRC for the same objects.

The conclusions from this work are as follows:

\begin{itemize}

\item We carried out a number of regression analyses in order to quantify the relation between $L_{150}$, SFR and mass for SFGs classified using BPT diagrams. While SFR is the dominant controlling parameter for radio luminosity of SFGs, as expected from earlier studies, both our stacking analysis and our multidimensional fitting to the data suggest a significant role for stellar mass as well.

\item We find that low-luminosity SFGs have a different relation between $L_{150}$ and SFR than the main SFG locus. Including stellar mass in the analysis reduces the scatter and allows us to demonstrate continuity between SFG and other objects, but does not change the basic picture in which low-luminosity sources are different from more luminous ones.  Simple models in which the calorimeter approximation breaks down in small, low-mass galaxies would predict that the radio luminosity of these systems would be smaller than the extrapolation from high-mass, high-SFR systems, but this is not the case. We suggest that at very low SFR we may be seeing an alternative mechanism for the generation of the radio-emitting cosmic rays such as pulsars or type Ia supernovae. Alternatively, amplification of magnetic fields in the high-SFR galaxies due to different galaxy dynamos or galactic winds could play a role. In our future work we will investigate the redshift evolution of the radio luminosity--SFR relation using deep LOFAR observations over the equatorial GAMA fields.

\item At least some of the objects unclassified on a BPT diagram that show excess radio emission are likely to be contaminated by low-level radio-loud AGN activity, as proposed by \cite{Hodge+08}.  However, the fact that many of them lie on the same mass-dependent sequence as low-SFR SFGs (Fig. \ref{l150-mass}) suggests that AGN contamination is far from universal. In our future work we will also study the nature of these objects in detail.

\item We find a tight relationship between the 150-MHz luminosity and the 250-$\mu$m far-infrared luminosity for our SFG samples. Comparison of the slopes obtained from the regression analyses (for both the FIRC and the radio luminosity--SFR relation) show that a flatter relation is found for $L_{1.4}$ than we obtain for $L_{150}$. This has been observed before and is explained by the effects of thermal radiation becoming more important at the higher radio frequency.

\end{itemize}

Sensitive studies of large numbers of galaxies in the nearby universe are crucial to establish well-calibrated relationships between radio luminosity, SFR and other galaxy parameters such as mass, which can then be used to probe SFR in the more distant universe where multiwavelength data are more limited.  The wide-area, sensitive LOFAR survey of the northern sky \citep{Shimwell+16}, particularly in combination with spectroscopic followup with WEAVE-LOFAR \citep{Smith+16}, will be key to establishing these relationships and will set the baseline for radio-based studies of star formation in the distant universe with the Square Kilometer Array in the coming years.


\section*{Acknowledgements}

 GG thanks the University of Hertfordshire for a PhD studentship. We thank the referee for her/his constructive comments. We would like to thank Mark Sargent for kindly providing the survival analysis tool and Gianfranco De Zotti for his useful comments. MJH and WLW acknowledge support from the UK Science and Technology Facilities Council [ST/M001008/1]. PNB is grateful for support from STFC via grant ST/M001229/1. HJAR and GCR gratefully acknowledge support from the European Research Council under the European Unions Seventh Framework Programme (FP/2007-2013)/ERC Advanced Grant NEWCLUSTERS-321271. JS is grateful for support from the UK STFC via grant ST/M001229/1. TS acknowledges support from the ERC Advanced Investigator programme NewClusters 321271. This research has made use of the University of Hertfordshire high-performance computing facility (\url{http://stri-cluster.herts.ac.uk/}) and the LOFAR-UK computing facility located at the University of Hertfordshire and supported by STFC [ST/P000096/1]. This research made use of {\sc astropy}, a community-developed core Python package for astronomy \citep{2013lf47} hosted at \url{http://www.astropy.org/} and of {\sc topcat} \citep{2005lf46}.

\textit{Herschel}-ATLAS is a project with \textit{Herschel}, which is an ESA space observatory with science instruments provided by European-led Principal Investigator consortia and with important participation from NASA. The H-ATLAS website is \url{http://www.h-atlas.org/}.

LOFAR, the Low Frequency Array designed and constructed by ASTRON, has facilities in several countries, that are owned by various parties (each with their own funding sources), and that are collectively operated by the International LOFAR Telescope (ILT) foundation under a joint scientific policy.

Funding for SDSS-III has been provided by the Alfred P. Sloan Foundation, the Participating Institutions, the National Science Foundation, and the U.S. Department of Energy Office of Science. The SDSS-III web site is \url{http://www.sdss3.org/}.

SDSS-III is managed by the Astrophysical Research Consortium for the Participating Institutions of the SDSS-III Collaboration including the University of Arizona, the Brazilian Participation Group, Brookhaven National Laboratory, Carnegie Mellon University, University of Florida, the French Participation Group, the German Participation Group, Harvard University, the Instituto de Astrofisica de Canarias, the Michigan State/Notre Dame/JINA Participation Group, Johns Hopkins University, Lawrence Berkeley National Laboratory, Max Planck Institute for Astrophysics, Max Planck Institute for Extraterrestrial Physics, New Mexico State University, New York University, Ohio State University, Pennsylvania State University, University of Portsmouth, Princeton University, the Spanish Participation Group, University of Tokyo, University of Utah, Vanderbilt University, University of Virginia, University of Washington, and Yale University.

The National Radio Astronomy Observatory (NRAO) is a facility of the National Science Foundation operated under cooperative agreement by Associated Universities, Inc.

\bibliographystyle{mnras}

\bibliography{ref3,ref3_,../bib/mjh,../bib/cards}

\begin{thebibliography}{}
\makeatletter
\relax
\def\mn@urlcharsother{\let\do\@makeother \do\$\do\&\do\#\do\^\do\_\do\%\do\~}
\def\mn@doi{\begingroup\mn@urlcharsother \@ifnextchar [ {\mn@doi@}
  {\mn@doi@[]}}
\def\mn@doi@[#1]#2{\def\@tempa{#1}\ifx\@tempa\@empty \href
  {http://dx.doi.org/#2} {doi:#2}\else \href {http://dx.doi.org/#2} {#1}\fi
  \endgroup}
\def\mn@eprint#1#2{\mn@eprint@#1:#2::\@nil}
\def\mn@eprint@arXiv#1{\href {http://arxiv.org/abs/#1} {{\tt arXiv:#1}}}
\def\mn@eprint@dblp#1{\href {http://dblp.uni-trier.de/rec/bibtex/#1.xml}
  {dblp:#1}}
\def\mn@eprint@#1:#2:#3:#4\@nil{\def\@tempa {#1}\def\@tempb {#2}\def\@tempc
  {#3}\ifx \@tempc \@empty \let \@tempc \@tempb \let \@tempb \@tempa \fi \ifx
  \@tempb \@empty \def\@tempb {arXiv}\fi \@ifundefined
  {mn@eprint@\@tempb}{\@tempb:\@tempc}{\expandafter \expandafter \csname
  mn@eprint@\@tempb\endcsname \expandafter{\@tempc}}}

\bibitem[\protect\citeauthoryear{{Abazajian} et~al.,}{{Abazajian}
  et~al.}{2009}]{2009ref113}
{Abazajian} K.~N.,  et~al., 2009, \mn@doi [\apjs]
  {10.1088/0067-0049/182/2/543}, \href
  {http://adsabs.harvard.edu/abs/2009ApJS..182..543A} {182, 543}

\bibitem[\protect\citeauthoryear{{Appleton} et~al.,}{{Appleton}
  et~al.}{2004}]{2004lf40}
{Appleton} P.~N.,  et~al., 2004, \mn@doi [\apjs] {10.1086/422425}, \href
  {http://adsabs.harvard.edu/abs/2004ApJS..154..147A} {154, 147}

\bibitem[\protect\citeauthoryear{{Astropy Collaboration} et~al.,}{{Astropy
  Collaboration} et~al.}{2013}]{2013lf47}
{Astropy Collaboration} et~al., 2013, \mn@doi [\aap]
  {10.1051/0004-6361/201322068}, \href
  {http://adsabs.harvard.edu/abs/2013A%26A...558A..33A} {558, A33}

\bibitem[\protect\citeauthoryear{{Baldwin}, {Phillips}  \&
  {Terlevich}}{{Baldwin} et~al.}{1981}]{1981ref115}
{Baldwin} J.~A.,  {Phillips} M.~M.,   {Terlevich} R.,  1981, \mn@doi [\pasp]
  {10.1086/130766}, \href {http://adsabs.harvard.edu/abs/1981PASP...93....5B}
  {93, 5}

\bibitem[\protect\citeauthoryear{{Becker}, {White}  \& {Helfand}}{{Becker}
  et~al.}{1995}]{1995ref118}
{Becker} R.~H.,  {White} R.~L.,   {Helfand} D.~J.,  1995, \mn@doi [\apj]
  {10.1086/176166}, \href {http://adsabs.harvard.edu/abs/1995ApJ...450..559B}
  {450, 559}

\bibitem[\protect\citeauthoryear{{Bell}}{{Bell}}{2003}]{2003lf10}
{Bell} E.~F.,  2003, \mn@doi [\apj] {10.1086/367829}, \href
  {http://adsabs.harvard.edu/abs/2003ApJ...586..794B} {586, 794}

\bibitem[\protect\citeauthoryear{{Berta} et~al.,}{{Berta}
  et~al.}{2013}]{berta13}
{Berta} S.,  et~al., 2013, \mn@doi [\aap] {10.1051/0004-6361/201220859}, \href
  {http://adsabs.harvard.edu/abs/2013A%26A...551A.100B} {551, A100}

\bibitem[\protect\citeauthoryear{{Best} \& {Heckman}}{{Best} \&
  {Heckman}}{2012}]{2012wref29}
{Best} P.~N.,  {Heckman} T.~M.,  2012, \mn@doi [\mnras]
  {10.1111/j.1365-2966.2012.20414.x}, \href
  {http://adsabs.harvard.edu/abs/2012MNRAS.421.1569B} {421, 1569}

\bibitem[\protect\citeauthoryear{{Best}, {Kauffmann}, {Heckman}, {Brinchmann},
  {Charlot}, {Ivezi{\'c}}  \& {White}}{{Best} et~al.}{2005}]{2005ref57}
{Best} P.~N.,  {Kauffmann} G.,  {Heckman} T.~M.,  {Brinchmann} J.,  {Charlot}
  S.,  {Ivezi{\'c}} {\v Z}.,   {White} S.~D.~M.,  2005, \mn@doi [\mnras]
  {10.1111/j.1365-2966.2005.09192.x}, \href
  {http://cdsads.u-strasbg.fr/abs/2005MNRAS.362...25B} {362, 25}

\bibitem[\protect\citeauthoryear{{Beswick}, {Muxlow}, {Thrall}, {Richards}  \&
  {Garrington}}{{Beswick} et~al.}{2008}]{2008lf11}
{Beswick} R.~J.,  {Muxlow} T.~W.~B.,  {Thrall} H.,  {Richards} A.~M.~S.,
  {Garrington} S.~T.,  2008, \mn@doi [\mnras]
  {10.1111/j.1365-2966.2008.12931.x}, \href
  {http://adsabs.harvard.edu/abs/2008MNRAS.385.1143B} {385, 1143}

\bibitem[\protect\citeauthoryear{{Bohlin}, {Dickinson}  \& {Calzetti}}{{Bohlin}
  et~al.}{2001}]{bohlin01}
{Bohlin} R.~C.,  {Dickinson} M.~E.,   {Calzetti} D.,  2001, \mn@doi [\aj]
  {10.1086/323137}, \href {http://adsabs.harvard.edu/abs/2001AJ....122.2118B}
  {122, 2118}

\bibitem[\protect\citeauthoryear{{Bonato} et~al.,}{{Bonato}
  et~al.}{2017}]{bonato17}
{Bonato} M.,  et~al., 2017, preprint, \href
  {http://adsabs.harvard.edu/abs/2017arXiv170405459B} {} (\mn@eprint {arXiv}
  {1704.05459})

\bibitem[\protect\citeauthoryear{{Bourne}, {Dunne}, {Ivison}, {Maddox},
  {Dickinson}  \& {Frayer}}{{Bourne} et~al.}{2011}]{2011lf24}
{Bourne} N.,  {Dunne} L.,  {Ivison} R.~J.,  {Maddox} S.~J.,  {Dickinson} M.,
  {Frayer} D.~T.,  2011, \mn@doi [\mnras] {10.1111/j.1365-2966.2010.17517.x},
  \href {http://adsabs.harvard.edu/abs/2011MNRAS.410.1155B} {410, 1155}

\bibitem[\protect\citeauthoryear{{Bourne} et~al.,}{{Bourne}
  et~al.}{2012a}]{Bourne2012}
{Bourne} N.,  et~al., 2012a, \mn@doi [\mnras]
  {10.1111/j.1365-2966.2012.20528.x}, \href
  {http://adsabs.harvard.edu/abs/2012MNRAS.421.3027B} {421, 3027}

\bibitem[\protect\citeauthoryear{{Bourne} et~al.,}{{Bourne}
  et~al.}{2012b}]{2012ref168}
{Bourne} N.,  et~al., 2012b, \mn@doi [\mnras]
  {10.1111/j.1365-2966.2012.20528.x}, \href
  {http://cdsads.u-strasbg.fr/abs/2012MNRAS.421.3027B} {421, 3027}

\bibitem[\protect\citeauthoryear{{Boyle}, {Cornwell}, {Middelberg}, {Norris},
  {Appleton}  \& {Smail}}{{Boyle} et~al.}{2007}]{2007intr161}
{Boyle} B.~J.,  {Cornwell} T.~J.,  {Middelberg} E.,  {Norris} R.~P.,
  {Appleton} P.~N.,   {Smail} I.,  2007, \mn@doi [\mnras]
  {10.1111/j.1365-2966.2007.11509.x}, \href
  {http://adsabs.harvard.edu/abs/2007MNRAS.376.1182B} {376, 1182}

\bibitem[\protect\citeauthoryear{{Brinchmann}, {Charlot}, {White}, {Tremonti},
  {Kauffmann}, {Heckman}  \& {Brinkmann}}{{Brinchmann} et~al.}{2004}]{2004ref8}
{Brinchmann} J.,  {Charlot} S.,  {White} S.~D.~M.,  {Tremonti} C.,  {Kauffmann}
  G.,  {Heckman} T.,   {Brinkmann} J.,  2004, \mn@doi [\mnras]
  {10.1111/j.1365-2966.2004.07881.x}, \href
  {http://cdsads.u-strasbg.fr/abs/2004MNRAS.351.1151B} {351, 1151}

\bibitem[\protect\citeauthoryear{{Brown} et~al.,}{{Brown}
  et~al.}{2014}]{brown14}
{Brown} M.~J.~I.,  et~al., 2014, \mn@doi [\apjs] {10.1088/0067-0049/212/2/18},
  \href {http://adsabs.harvard.edu/abs/2014ApJS..212...18B} {212, 18}

\bibitem[\protect\citeauthoryear{{Brown} et~al.,}{{Brown}
  et~al.}{2017}]{brown17}
{Brown} M.~J.~I.,  et~al., 2017, \mn@doi [\apj] {10.3847/1538-4357/aa8ad2},
  \href {http://adsabs.harvard.edu/abs/2017ApJ...847..136B} {847, 136}

\bibitem[\protect\citeauthoryear{{Bruzual} \& {Charlot}}{{Bruzual} \&
  {Charlot}}{2003}]{bc03}
{Bruzual} G.,  {Charlot} S.,  2003, \mn@doi [\mnras]
  {10.1046/j.1365-8711.2003.06897.x}, \href
  {http://adsabs.harvard.edu/abs/2003MNRAS.344.1000B} {344, 1000}

\bibitem[\protect\citeauthoryear{{Calistro Rivera} et~al.,}{{Calistro Rivera}
  et~al.}{2017}]{Calistro17}
{Calistro Rivera} G.,  et~al., 2017, preprint, \href
  {http://adsabs.harvard.edu/abs/2017arXiv170406268C} {} (\mn@eprint {arXiv}
  {1704.06268})

\bibitem[\protect\citeauthoryear{{Calzetti}}{{Calzetti}}{2013}]{2013lfr1}
{Calzetti} D.,  2013, {Star Formation Rate Indicators}.
p.~419

\bibitem[\protect\citeauthoryear{{Casey} et~al.,}{{Casey}
  et~al.}{2014}]{Casey2014}
{Casey} C.~M.,  et~al., 2014, \mn@doi [\apj] {10.1088/0004-637X/796/2/95},
  \href {http://adsabs.harvard.edu/abs/2014ApJ...796...95C} {796, 95}

\bibitem[\protect\citeauthoryear{{Chabrier}}{{Chabrier}}{2003}]{Chabrier03}
{Chabrier} G.,  2003, \mn@doi [\apjl] {10.1086/374879}, \href
  {http://adsabs.harvard.edu/abs/2003ApJ...586L.133C} {586, L133}

\bibitem[\protect\citeauthoryear{{Charlot} \& {Fall}}{{Charlot} \&
  {Fall}}{2000}]{charlot00}
{Charlot} S.,  {Fall} S.~M.,  2000, \mn@doi [\apj] {10.1086/309250}, \href
  {http://adsabs.harvard.edu/abs/2000ApJ...539..718C} {539, 718}

\bibitem[\protect\citeauthoryear{{Chi} \& {Wolfendale}}{{Chi} \&
  {Wolfendale}}{1990}]{1990chi}
{Chi} X.,  {Wolfendale} A.~W.,  1990, \mnras, \href
  {http://adsabs.harvard.edu/abs/1990MNRAS.245..101C} {245, 101}

\bibitem[\protect\citeauthoryear{{Condon}}{{Condon}}{1992}]{1992lf5}
{Condon} J.~J.,  1992, \mn@doi [\araa] {10.1146/annurev.aa.30.090192.003043},
  \href {http://cdsads.u-strasbg.fr/abs/1992ARA%26A..30..575C} {30, 575}

\bibitem[\protect\citeauthoryear{{Condon}, {Cotton}, {Greisen}, {Yin},
  {Perley}, {Taylor}  \& {Broderick}}{{Condon} et~al.}{1998}]{1998ref117}
{Condon} J.~J.,  {Cotton} W.~D.,  {Greisen} E.~W.,  {Yin} Q.~F.,  {Perley}
  R.~A.,  {Taylor} G.~B.,   {Broderick} J.~J.,  1998, \mn@doi [\aj]
  {10.1086/300337}, \href {http://cdsads.u-strasbg.fr/abs/1998AJ....115.1693C}
  {115, 1693}

\bibitem[\protect\citeauthoryear{{Cox}, {Eales}, {Alexander}  \& {Fitt}}{{Cox}
  et~al.}{1988}]{1988lf52}
{Cox} M.~J.,  {Eales} S.~A.~E.,  {Alexander} P.,   {Fitt} A.~J.,  1988, \mn@doi
  [\mnras] {10.1093/mnras/235.4.1227}, \href
  {http://adsabs.harvard.edu/abs/1988MNRAS.235.1227C} {235, 1227}

\bibitem[\protect\citeauthoryear{{Cram}, {Hopkins}, {Mobasher}  \&
  {Rowan-Robinson}}{{Cram} et~al.}{1998}]{Cram+98}
{Cram} L.,  {Hopkins} A.,  {Mobasher} B.,   {Rowan-Robinson} M.,  1998, \mn@doi
  [\apj] {10.1086/306333}, \href
  {http://adsabs.harvard.edu/abs/1998ApJ...507..155C} {507, 155}

\bibitem[\protect\citeauthoryear{{Dariush} et~al.,}{{Dariush}
  et~al.}{2016}]{dariush16}
{Dariush} A.,  et~al., 2016, \mn@doi [\mnras] {10.1093/mnras/stv2767}, \href
  {http://adsabs.harvard.edu/abs/2016MNRAS.456.2221D} {456, 2221}

\bibitem[\protect\citeauthoryear{{Davies} et~al.,}{{Davies}
  et~al.}{2016}]{2016davies}
{Davies} L.~J.~M.,  et~al., 2016, \mn@doi [\mnras] {10.1093/mnras/stw1342},
  \href {http://adsabs.harvard.edu/abs/2016MNRAS.461..458D} {461, 458}

\bibitem[\protect\citeauthoryear{{Davies} et~al.,}{{Davies}
  et~al.}{2017}]{davies+17}
{Davies} L.~J.~M.,  et~al., 2017, \mn@doi [\mnras] {10.1093/mnras/stw3080},
  \href {http://adsabs.harvard.edu/abs/2017MNRAS.466.2312D} {466, 2312}

\bibitem[\protect\citeauthoryear{{Delhaize} et~al.,}{{Delhaize}
  et~al.}{2017}]{delhaize17}
{Delhaize} J.,  et~al., 2017, preprint, \href
  {http://adsabs.harvard.edu/abs/2017arXiv170309723D} {} (\mn@eprint {arXiv}
  {1703.09723})

\bibitem[\protect\citeauthoryear{{Donoso}, {Best}  \& {Kauffmann}}{{Donoso}
  et~al.}{2009}]{2009ref169}
{Donoso} E.,  {Best} P.~N.,   {Kauffmann} G.,  2009, \mn@doi [\mnras]
  {10.1111/j.1365-2966.2008.14068.x}, \href
  {http://adsabs.harvard.edu/abs/2009MNRAS.392..617D} {392, 617}

\bibitem[\protect\citeauthoryear{{Eales} et~al.,}{{Eales}
  et~al.}{2010}]{2010ref1}
{Eales} S.,  et~al., 2010, \mn@doi [\pasp] {10.1086/653086}, \href
  {http://esoads.eso.org/abs/2010PASP..122..499E} {122, 499}

\bibitem[\protect\citeauthoryear{{Eales} et~al.,}{{Eales}
  et~al.}{2015}]{eales15}
{Eales} S.,  et~al., 2015, \mn@doi [\mnras] {10.1093/mnras/stv1300}, \href
  {http://adsabs.harvard.edu/abs/2015MNRAS.452.3489E} {452, 3489}

\bibitem[\protect\citeauthoryear{{Elbaz} et~al.,}{{Elbaz}
  et~al.}{2007a}]{2007lf49}
{Elbaz} D.,  et~al., 2007a, \mn@doi [\aap] {10.1051/0004-6361:20077525}, \href
  {http://adsabs.harvard.edu/abs/2007A%26A...468...33E} {468, 33}

\bibitem[\protect\citeauthoryear{{Elbaz} et~al.,}{{Elbaz}
  et~al.}{2007b}]{Elbaz07}
{Elbaz} D.,  et~al., 2007b, \mn@doi [\aap] {10.1051/0004-6361:20077525}, \href
  {http://adsabs.harvard.edu/abs/2007A%26A...468...33E} {468, 33}

\bibitem[\protect\citeauthoryear{{Foreman-Mackey}, {Hogg}, {Lang}  \&
  {Goodman}}{{Foreman-Mackey} et~al.}{2013}]{Foreman-Mackey+13}
{Foreman-Mackey} D.,  {Hogg} D.~W.,  {Lang} D.,   {Goodman} J.,  2013, \mn@doi
  [\pasp] {10.1086/670067}, \href
  {http://adsabs.harvard.edu/abs/2013PASP..125..306F} {125, 306}

\bibitem[\protect\citeauthoryear{{Garn}, {Green}, {Riley}  \&
  {Alexander}}{{Garn} et~al.}{2009}]{2009lf43}
{Garn} T.,  {Green} D.~A.,  {Riley} J.~M.,   {Alexander} P.,  2009, \mn@doi
  [\mnras] {10.1111/j.1365-2966.2009.15073.x}, \href
  {http://adsabs.harvard.edu/abs/2009MNRAS.397.1101G} {397, 1101}

\bibitem[\protect\citeauthoryear{{Griffin} et~al.,}{{Griffin}
  et~al.}{2010}]{2010ref11}
{Griffin} M.~J.,  et~al., 2010, \mn@doi [\aap] {10.1051/0004-6361/201014519},
  \href {http://esoads.eso.org/abs/2010A%26A...518L...3G} {518, L3}

\bibitem[\protect\citeauthoryear{{G{\"u}rkan} et~al.,}{{G{\"u}rkan}
  et~al.}{2015}]{2015lf17}
{G{\"u}rkan} G.,  et~al., 2015, \mn@doi [\mnras] {10.1093/mnras/stv1502}, \href
  {http://adsabs.harvard.edu/abs/2015MNRAS.452.3776G} {452, 3776}

\bibitem[\protect\citeauthoryear{{Hardcastle} et~al.,}{{Hardcastle}
  et~al.}{2010}]{2010ref7}
{Hardcastle} M.~J.,  et~al., 2010, \mn@doi [\mnras]
  {10.1111/j.1365-2966.2010.17791.x}, \href
  {http://cdsads.u-strasbg.fr/abs/2010MNRAS.409..122H} {409, 122}

\bibitem[\protect\citeauthoryear{{Hardcastle} et~al.,}{{Hardcastle}
  et~al.}{2013}]{2013ref9}
{Hardcastle} M.~J.,  et~al., 2013, \mn@doi [\mnras] {10.1093/mnras/sts510},
  \href {http://esoads.eso.org/abs/2013MNRAS.429.2407H} {429, 2407}

\bibitem[\protect\citeauthoryear{{Hardcastle} et~al.,}{{Hardcastle}
  et~al.}{2016}]{Hardcastle+16}
{Hardcastle} M.~J.,  et~al., 2016, \mn@doi [\mnras] {10.1093/mnras/stw1763},
  \href {http://adsabs.harvard.edu/abs/2016MNRAS.462.1910H} {462, 1910}

\bibitem[\protect\citeauthoryear{{Harwit} \& {Pacini}}{{Harwit} \&
  {Pacini}}{1975}]{1975lf7}
{Harwit} M.,  {Pacini} F.,  1975, \mn@doi [\apjl] {10.1086/181913}, \href
  {http://cdsads.u-strasbg.fr/abs/1975ApJ...200L.127H} {200, L127}

\bibitem[\protect\citeauthoryear{{Hayward} \& {Smith}}{{Hayward} \&
  {Smith}}{2015}]{hayward15}
{Hayward} C.~C.,  {Smith} D.~J.~B.,  2015, \mn@doi [\mnras]
  {10.1093/mnras/stu2195}, \href
  {http://adsabs.harvard.edu/abs/2015MNRAS.446.1512H} {446, 1512}

\bibitem[\protect\citeauthoryear{{Hayward}, {Jonsson}, {Kere{\v s}},
  {Magnelli}, {Hernquist}  \& {Cox}}{{Hayward} et~al.}{2012}]{hayward12}
{Hayward} C.~C.,  {Jonsson} P.,  {Kere{\v s}} D.,  {Magnelli} B.,  {Hernquist}
  L.,   {Cox} T.~J.,  2012, \mn@doi [\mnras]
  {10.1111/j.1365-2966.2012.21254.x}, \href
  {http://adsabs.harvard.edu/abs/2012MNRAS.424..951H} {424, 951}

\bibitem[\protect\citeauthoryear{{Heesen}, {Beck}, {Krause}  \&
  {Dettmar}}{{Heesen} et~al.}{2009}]{2009heesen}
{Heesen} V.,  {Beck} R.,  {Krause} M.,   {Dettmar} R.-J.,  2009, \mn@doi
  [Astronomische Nachrichten] {10.1002/asna.200911279}, \href
  {http://adsabs.harvard.edu/abs/2009AN....330.1028H} {330, 1028}

\bibitem[\protect\citeauthoryear{{Helou} \& {Bicay}}{{Helou} \&
  {Bicay}}{1993}]{1993lf36}
{Helou} G.,  {Bicay} M.~D.,  1993, \mn@doi [\apj] {10.1086/173146}, \href
  {http://adsabs.harvard.edu/abs/1993ApJ...415...93H} {415, 93}

\bibitem[\protect\citeauthoryear{{Helou}, {Soifer}  \&
  {Rowan-Robinson}}{{Helou} et~al.}{1985}]{Helou+85}
{Helou} G.,  {Soifer} B.~T.,   {Rowan-Robinson} M.,  1985, \mn@doi [\apjl]
  {10.1086/184556}, \href {http://cdsads.u-strasbg.fr/abs/1985ApJ...298L...7H}
  {298, L7}

\bibitem[\protect\citeauthoryear{{Hildebrand}}{{Hildebrand}}{1983}]{hildebrand83}
{Hildebrand} R.~H.,  1983, \qjras, \href
  {http://adsabs.harvard.edu/abs/1983QJRAS..24..267H} {24, 267}

\bibitem[\protect\citeauthoryear{{Hodge}, {Becker}, {White}  \& {de
  Vries}}{{Hodge} et~al.}{2008}]{Hodge+08}
{Hodge} J.~A.,  {Becker} R.~H.,  {White} R.~L.,   {de Vries} W.~H.,  2008,
  \mn@doi [\aj] {10.1088/0004-6256/136/3/1097}, \href
  {http://adsabs.harvard.edu/abs/2008AJ....136.1097H} {136, 1097}

\bibitem[\protect\citeauthoryear{{Hummel}, {Davies}, {Pedlar}, {Wolstencroft}
  \& {van der Hulst}}{{Hummel} et~al.}{1988}]{1988lf6}
{Hummel} E.,  {Davies} R.~D.,  {Pedlar} A.,  {Wolstencroft} R.~D.,   {van der
  Hulst} J.~M.,  1988, \aap, \href
  {http://cdsads.u-strasbg.fr/abs/1988A%26A...199...91H} {199, 91}

\bibitem[\protect\citeauthoryear{{Ibar} et~al.,}{{Ibar}
  et~al.}{2010}]{2010ref10}
{Ibar} E.,  et~al., 2010, \mn@doi [\mnras] {10.1111/j.1365-2966.2010.17620.x},
  \href {http://esoads.eso.org/abs/2010MNRAS.409...38I} {409, 38}

\bibitem[\protect\citeauthoryear{{Israel}, {Mahoney}  \& {Howarth}}{{Israel}
  et~al.}{1992}]{israel92}
{Israel} F.~P.,  {Mahoney} M.~J.,   {Howarth} N.,  1992, \aap, \href
  {http://adsabs.harvard.edu/abs/1992A%26A...261...47I} {261, 47}

\bibitem[\protect\citeauthoryear{{Ivison} et~al.,}{{Ivison}
  et~al.}{2010a}]{2010lf29}
{Ivison} R.~J.,  et~al., 2010a, \mn@doi [\mnras]
  {10.1111/j.1365-2966.2009.15918.x}, \href
  {http://adsabs.harvard.edu/abs/2010MNRAS.402..245I} {402, 245}

\bibitem[\protect\citeauthoryear{{Ivison} et~al.,}{{Ivison}
  et~al.}{2010b}]{Ivison+10}
{Ivison} R.~J.,  et~al., 2010b, \mn@doi [\aap] {10.1051/0004-6361/201014552},
  \href {http://adsabs.harvard.edu/abs/2010A%26A...518L..31I} {518, L31}

\bibitem[\protect\citeauthoryear{{Jarvis} et~al.,}{{Jarvis}
  et~al.}{2010}]{Jarvis+10}
{Jarvis} M.~J.,  et~al., 2010, \mn@doi [\mnras]
  {10.1111/j.1365-2966.2010.17772.x}, \href
  {http://esoads.eso.org/abs/2010MNRAS.409...92J} {409, 92}

\bibitem[\protect\citeauthoryear{{Johnston}, {Vaccari}, {Jarvis}, {Smith},
  {Giovannoli}, {H{\"a}u{\ss}ler}  \& {Prescott}}{{Johnston}
  et~al.}{2015}]{Johnston2015}
{Johnston} R.,  {Vaccari} M.,  {Jarvis} M.,  {Smith} M.,  {Giovannoli} E.,
  {H{\"a}u{\ss}ler} B.,   {Prescott} M.,  2015, \mn@doi [\mnras]
  {10.1093/mnras/stv1715}, \href
  {http://adsabs.harvard.edu/abs/2015MNRAS.453.2540J} {453, 2540}

\bibitem[\protect\citeauthoryear{{Kapi{\'n}ska} et~al.,}{{Kapi{\'n}ska}
  et~al.}{2017}]{kapinska17}
{Kapi{\'n}ska} A.~D.,  et~al., 2017, \mn@doi [\apj] {10.3847/1538-4357/aa5f5d},
  \href {http://adsabs.harvard.edu/abs/2017ApJ...838...68K} {838, 68}

\bibitem[\protect\citeauthoryear{{Karim} et~al.,}{{Karim}
  et~al.}{2011}]{2011lf50}
{Karim} A.,  et~al., 2011, \mn@doi [\apj] {10.1088/0004-637X/730/2/61}, \href
  {http://adsabs.harvard.edu/abs/2011ApJ...730...61K} {730, 61}

\bibitem[\protect\citeauthoryear{{Kauffmann} et~al.,}{{Kauffmann}
  et~al.}{2003}]{2003ref88}
{Kauffmann} G.,  et~al., 2003, \mn@doi [\mnras]
  {10.1111/j.1365-2966.2003.07154.x}, \href
  {http://cdsads.u-strasbg.fr/abs/2003MNRAS.346.1055K} {346, 1055}

\bibitem[\protect\citeauthoryear{{Kennicutt} \& {Evans}}{{Kennicutt} \&
  {Evans}}{2012}]{2012lfr2}
{Kennicutt} R.~C.,  {Evans} N.~J.,  2012, \mn@doi [\araa]
  {10.1146/annurev-astro-081811-125610}, \href
  {http://adsabs.harvard.edu/abs/2012ARA%26A..50..531K} {50, 531}

\bibitem[\protect\citeauthoryear{{Kewley}, {Groves}, {Kauffmann}  \&
  {Heckman}}{{Kewley} et~al.}{2006}]{2006ref2}
{Kewley} L.~J.,  {Groves} B.,  {Kauffmann} G.,   {Heckman} T.,  2006, \mn@doi
  [\mnras] {10.1111/j.1365-2966.2006.10859.x}, \href
  {http://esoads.eso.org/abs/2006MNRAS.372..961K} {372, 961}

\bibitem[\protect\citeauthoryear{{Klein}}{{Klein}}{1991}]{1991lf45}
{Klein} U.,  1991, Proceedings of the Astronomical Society of Australia, \href
  {http://adsabs.harvard.edu/abs/1991PASAu...9..253K} {9, 253}

\bibitem[\protect\citeauthoryear{{Lacki} \& {Thompson}}{{Lacki} \&
  {Thompson}}{2010}]{2010lf20}
{Lacki} B.~C.,  {Thompson} T.~A.,  2010, \mn@doi [\apj]
  {10.1088/0004-637X/717/1/196}, \href
  {http://adsabs.harvard.edu/abs/2010ApJ...717..196L} {717, 196}

\bibitem[\protect\citeauthoryear{{Lacki}, {Thompson}  \& {Quataert}}{{Lacki}
  et~al.}{2010}]{2010lf19}
{Lacki} B.~C.,  {Thompson} T.~A.,   {Quataert} E.,  2010, \mn@doi [\apj]
  {10.1088/0004-637X/717/1/1}, \href
  {http://adsabs.harvard.edu/abs/2010ApJ...717....1L} {717, 1}

\bibitem[\protect\citeauthoryear{{Lang}}{{Lang}}{2014}]{lang14a}
{Lang} D.,  2014, \mn@doi [\aj] {10.1088/0004-6256/147/5/108}, \href
  {http://adsabs.harvard.edu/abs/2014AJ....147..108L} {147, 108}

\bibitem[\protect\citeauthoryear{{Lang}, {Hogg}  \& {Schlegel}}{{Lang}
  et~al.}{2014}]{lang14b}
{Lang} D.,  {Hogg} D.~W.,   {Schlegel} D.~J.,  2014, preprint, \href
  {http://adsabs.harvard.edu/abs/2014arXiv1410.7397L} {} (\mn@eprint {arXiv}
  {1410.7397})

\bibitem[\protect\citeauthoryear{{Lange} et~al.,}{{Lange}
  et~al.}{2015}]{2015lange}
{Lange} R.,  et~al., 2015, \mn@doi [\mnras] {10.1093/mnras/stu2467}, \href
  {http://adsabs.harvard.edu/abs/2015MNRAS.447.2603L} {447, 2603}

\bibitem[\protect\citeauthoryear{{Lanz} et~al.,}{{Lanz} et~al.}{2013}]{lanz13}
{Lanz} L.,  et~al., 2013, \mn@doi [\apj] {10.1088/0004-637X/768/1/90}, \href
  {http://adsabs.harvard.edu/abs/2013ApJ...768...90L} {768, 90}

\bibitem[\protect\citeauthoryear{{Leslie}, {Kewley}, {Sanders}  \&
  {Lee}}{{Leslie} et~al.}{2016}]{leslie16}
{Leslie} S.~K.,  {Kewley} L.~J.,  {Sanders} D.~B.,   {Lee} N.,  2016, \mn@doi
  [\mnras] {10.1093/mnrasl/slv135}, \href
  {http://adsabs.harvard.edu/abs/2016MNRAS.455L..82L} {455, L82}

\bibitem[\protect\citeauthoryear{{Lisenfeld}, {Voelk}  \& {Xu}}{{Lisenfeld}
  et~al.}{1996}]{1996lf34}
{Lisenfeld} U.,  {Voelk} H.~J.,   {Xu} C.,  1996, \aap, \href
  {http://adsabs.harvard.edu/abs/1996A%26A...306..677L} {306, 677}

\bibitem[\protect\citeauthoryear{{Magnelli} et~al.,}{{Magnelli}
  et~al.}{2015}]{2015lf30}
{Magnelli} B.,  et~al., 2015, \mn@doi [\aap] {10.1051/0004-6361/201424937},
  \href {http://adsabs.harvard.edu/abs/2015A%26A...573A..45M} {573, A45}

\bibitem[\protect\citeauthoryear{{Mancuso} et~al.,}{{Mancuso}
  et~al.}{2015}]{Mancuso15}
{Mancuso} C.,  et~al., 2015, \mn@doi [\apj] {10.1088/0004-637X/810/1/72}, \href
  {http://adsabs.harvard.edu/abs/2015ApJ...810...72M} {810, 72}

\bibitem[\protect\citeauthoryear{{Martin} et~al.,}{{Martin}
  et~al.}{2005}]{2005lf33}
{Martin} D.~C.,  et~al., 2005, \mn@doi [\apjl] {10.1086/425496}, \href
  {http://adsabs.harvard.edu/abs/2005ApJ...619L..59M} {619, L59}

\bibitem[\protect\citeauthoryear{{Massardi}, {Bonaldi}, {Negrello},
  {Ricciardi}, {Raccanelli}  \& {de Zotti}}{{Massardi}
  et~al.}{2010}]{massardi10}
{Massardi} M.,  {Bonaldi} A.,  {Negrello} M.,  {Ricciardi} S.,  {Raccanelli}
  A.,   {de Zotti} G.,  2010, \mn@doi [\mnras]
  {10.1111/j.1365-2966.2010.16305.x}, \href
  {http://adsabs.harvard.edu/abs/2010MNRAS.404..532M} {404, 532}

\bibitem[\protect\citeauthoryear{{Mohan} \& {Rafferty}}{{Mohan} \&
  {Rafferty}}{2015}]{Mohan+Rafferty15}
{Mohan} N.,  {Rafferty} D.,  2015, {PyBDSF: Python Blob Detection and Source
  Finder}, Astrophysics Source Code Library (\mn@eprint {ascl} {1502.007})

\bibitem[\protect\citeauthoryear{{Murphy} et~al.,}{{Murphy}
  et~al.}{2011}]{murphy+11}
{Murphy} E.~J.,  et~al., 2011, \mn@doi [\apj] {10.1088/0004-637X/737/2/67},
  \href {http://adsabs.harvard.edu/abs/2011ApJ...737...67M} {737, 67}

\bibitem[\protect\citeauthoryear{{Negrello} et~al.,}{{Negrello}
  et~al.}{2014}]{negrello14}
{Negrello} M.,  et~al., 2014, \mn@doi [\mnras] {10.1093/mnras/stu413}, \href
  {http://adsabs.harvard.edu/abs/2014MNRAS.440.1999N} {440, 1999}

\bibitem[\protect\citeauthoryear{{Niklas}}{{Niklas}}{1997}]{1997lf42}
{Niklas} S.,  1997, \aap, \href
  {http://adsabs.harvard.edu/abs/1997A%26A...322...29N} {322, 29}

\bibitem[\protect\citeauthoryear{{Niklas} \& {Beck}}{{Niklas} \&
  {Beck}}{1997}]{1997lf37}
{Niklas} S.,  {Beck} R.,  1997, \aap, \href
  {http://adsabs.harvard.edu/abs/1997A%26A...320...54N} {320, 54}

\bibitem[\protect\citeauthoryear{{Noeske} et~al.,}{{Noeske}
  et~al.}{2007}]{2007lf48}
{Noeske} K.~G.,  et~al., 2007, \mn@doi [\apjl] {10.1086/517926}, \href
  {http://adsabs.harvard.edu/abs/2007ApJ...660L..43N} {660, L43}

\bibitem[\protect\citeauthoryear{{Oke} \& {Gunn}}{{Oke} \&
  {Gunn}}{1983}]{oke83}
{Oke} J.~B.,  {Gunn} J.~E.,  1983, \mn@doi [\apj] {10.1086/160817}, \href
  {http://adsabs.harvard.edu/abs/1983ApJ...266..713O} {266, 713}

\bibitem[\protect\citeauthoryear{{Overzier} et~al.,}{{Overzier}
  et~al.}{2011}]{Overzier2011}
{Overzier} R.~A.,  et~al., 2011, \mn@doi [\apjl] {10.1088/2041-8205/726/1/L7},
  \href {http://adsabs.harvard.edu/abs/2011ApJ...726L...7O} {726, L7}

\bibitem[\protect\citeauthoryear{{Pascale} et~al.,}{{Pascale}
  et~al.}{2011}]{2011ref145}
{Pascale} E.,  et~al., 2011, \mn@doi [\mnras]
  {10.1111/j.1365-2966.2011.18756.x}, \href
  {http://esoads.eso.org/abs/2011MNRAS.415..911P} {415, 911}

\bibitem[\protect\citeauthoryear{{Poglitsch} et~al.,}{{Poglitsch}
  et~al.}{2010}]{poglitsch10}
{Poglitsch} A.,  et~al., 2010, \mn@doi [\aap] {10.1051/0004-6361/201014535},
  \href {http://adsabs.harvard.edu/abs/2010A%26A...518L...2P} {518, L2}

\bibitem[\protect\citeauthoryear{{Price} \& {Duric}}{{Price} \&
  {Duric}}{1992}]{1992lf41}
{Price} R.,  {Duric} N.,  1992, \mn@doi [\apj] {10.1086/172040}, \href
  {http://adsabs.harvard.edu/abs/1992ApJ...401...81P} {401, 81}

\bibitem[\protect\citeauthoryear{{Rickard} \& {Harvey}}{{Rickard} \&
  {Harvey}}{1984}]{1984lf8}
{Rickard} L.~J.,  {Harvey} P.~M.,  1984, \mn@doi [\aj] {10.1086/113652}, \href
  {http://cdsads.u-strasbg.fr/abs/1984AJ.....89.1520R} {89, 1520}

\bibitem[\protect\citeauthoryear{{Rowlands} et~al.,}{{Rowlands}
  et~al.}{2014}]{rowlands14}
{Rowlands} K.,  et~al., 2014, \mn@doi [\mnras] {10.1093/mnras/stu510}, \href
  {http://adsabs.harvard.edu/abs/2014MNRAS.441.1017R} {441, 1017}

\bibitem[\protect\citeauthoryear{{Sargent} et~al.,}{{Sargent}
  et~al.}{2010}]{Sargent+10}
{Sargent} M.~T.,  et~al., 2010, \mn@doi [\apjs] {10.1088/0067-0049/186/2/341},
  \href {http://adsabs.harvard.edu/abs/2010ApJS..186..341S} {186, 341}

\bibitem[\protect\citeauthoryear{{Schlegel}, {Finkbeiner}  \&
  {Davis}}{{Schlegel} et~al.}{1998}]{schlegel98}
{Schlegel} D.~J.,  {Finkbeiner} D.~P.,   {Davis} M.,  1998, \mn@doi [\apj]
  {10.1086/305772}, \href {http://adsabs.harvard.edu/abs/1998ApJ...500..525S}
  {500, 525}

\bibitem[\protect\citeauthoryear{{Schleicher} \& {Beck}}{{Schleicher} \&
  {Beck}}{2013}]{Schleicher+Beck13}
{Schleicher} D.~R.~G.,  {Beck} R.,  2013, \mn@doi [\aap]
  {10.1051/0004-6361/201321707}, \href
  {http://adsabs.harvard.edu/abs/2013A%26A...556A.142S} {556, A142}

\bibitem[\protect\citeauthoryear{{Schmitt}, {Kahabka}, {Stauffer}  \&
  {Piters}}{{Schmitt} et~al.}{1993}]{Schmitt+93}
{Schmitt} J.~H.~M.~M.,  {Kahabka} P.,  {Stauffer} J.,   {Piters} A.~J.~M.,
  1993, \aap, \href {http://adsabs.harvard.edu/abs/1993A%26A...277..114S} {277,
  114}

\bibitem[\protect\citeauthoryear{{Schober}, {Schleicher}  \&
  {Klessen}}{{Schober} et~al.}{2017}]{schober+17}
{Schober} J.,  {Schleicher} D.~R.~G.,   {Klessen} R.~S.,  2017, \mn@doi
  [\mnras] {10.1093/mnras/stx460}, \href
  {http://adsabs.harvard.edu/abs/2017MNRAS.468..946S} {468, 946}

\bibitem[\protect\citeauthoryear{{Shimwell} et~al.,}{{Shimwell}
  et~al.}{2017}]{Shimwell+16}
{Shimwell} T.~W.,  et~al., 2017, \mn@doi [\aap] {10.1051/0004-6361/201629313},
  \href {http://adsabs.harvard.edu/abs/2017A%26A...598A.104S} {598, A104}

\bibitem[\protect\citeauthoryear{{Smith} \& {Hayward}}{{Smith} \&
  {Hayward}}{2015}]{smith15}
{Smith} D.~J.~B.,  {Hayward} C.~C.,  2015, \mn@doi [\mnras]
  {10.1093/mnras/stv1727}, \href
  {http://adsabs.harvard.edu/abs/2015MNRAS.453.1597S} {453, 1597}

\bibitem[\protect\citeauthoryear{{Smith} et~al.,}{{Smith}
  et~al.}{2012}]{smith12}
{Smith} D.~J.~B.,  et~al., 2012, \mn@doi [\mnras]
  {10.1111/j.1365-2966.2012.21930.x}, \href
  {http://adsabs.harvard.edu/abs/2012MNRAS.427..703S} {427, 703}

\bibitem[\protect\citeauthoryear{{Smith} et~al.,}{{Smith}
  et~al.}{2013}]{2013ref87}
{Smith} D.~J.~B.,  et~al., 2013, \mn@doi [\mnras] {10.1093/mnras/stt1737},
  \href {http://esoads.eso.org/abs/2013MNRAS.436.2435S} {436, 2435}

\bibitem[\protect\citeauthoryear{{Smith} et~al.,}{{Smith}
  et~al.}{2014}]{Smith+14}
{Smith} D.~J.~B.,  et~al., 2014, \mn@doi [\mnras] {10.1093/mnras/stu1830},
  \href {http://adsabs.harvard.edu/abs/2014MNRAS.445.2232S} {445, 2232}

\bibitem[\protect\citeauthoryear{{Smith} et~al.,}{{Smith}
  et~al.}{2016}]{Smith+16}
{Smith} D.~J.~B.,  et~al., 2016, preprint, \href
  {http://adsabs.harvard.edu/abs/2016arXiv161102706S} {} (\mn@eprint {arXiv}
  {1611.02706})

\bibitem[\protect\citeauthoryear{{Smol{\v c}i{\'c}}}{{Smol{\v
  c}i{\'c}}}{2009}]{smolcic+9}
{Smol{\v c}i{\'c}} V.,  2009, \mn@doi [\apjl] {10.1088/0004-637X/699/1/L43},
  \href {http://adsabs.harvard.edu/abs/2009ApJ...699L..43S} {699, L43}

\bibitem[\protect\citeauthoryear{{Tabatabaei} et~al.,}{{Tabatabaei}
  et~al.}{2017}]{taba+17}
{Tabatabaei} F.~S.,  et~al., 2017, \mn@doi [\apj]
  {10.3847/1538-4357/836/2/185}, \href
  {http://adsabs.harvard.edu/abs/2017ApJ...836..185T} {836, 185}

\bibitem[\protect\citeauthoryear{{Takeuchi}, {Yuan}, {Ikeyama}, {Murata}  \&
  {Inoue}}{{Takeuchi} et~al.}{2012}]{Takeuchi2012}
{Takeuchi} T.~T.,  {Yuan} F.-T.,  {Ikeyama} A.,  {Murata} K.~L.,   {Inoue}
  A.~K.,  2012, \mn@doi [\apj] {10.1088/0004-637X/755/2/144}, \href
  {http://adsabs.harvard.edu/abs/2012ApJ...755..144T} {755, 144}

\bibitem[\protect\citeauthoryear{{Tasse}}{{Tasse}}{2014a}]{Tasse14b}
{Tasse} C.,  2014a, preprint, \href
  {http://adsabs.harvard.edu/abs/2014arXiv1410.8706T} {} (\mn@eprint {arXiv}
  {1410.8706})

\bibitem[\protect\citeauthoryear{{Tasse}}{{Tasse}}{2014b}]{Tasse14a}
{Tasse} C.,  2014b, \mn@doi [\aap] {10.1051/0004-6361/201423503}, \href
  {http://adsabs.harvard.edu/abs/2014A%26A...566A.127T} {566, A127}

\bibitem[\protect\citeauthoryear{{Tasse} et~al.,}{{Tasse}
  et~al.}{2017}]{Tasse17}
{Tasse} C.,  et~al., 2017, preprint, \href
  {http://adsabs.harvard.edu/abs/2017arXiv171202078T} {} (\mn@eprint {arXiv}
  {1712.02078})

\bibitem[\protect\citeauthoryear{{Taylor}}{{Taylor}}{2005}]{2005lf46}
{Taylor} M.~B.,  2005, in {Shopbell} P.,  {Britton} M.,   {Ebert} R.,  eds,
  Astronomical Society of the Pacific Conference Series Vol. 347, Astronomical
  Data Analysis Software and Systems XIV. p.~29

\bibitem[\protect\citeauthoryear{{Valiante} et~al.,}{{Valiante}
  et~al.}{2016}]{2016valiante}
{Valiante} E.,  et~al., 2016, \mn@doi [\mnras] {10.1093/mnras/stw1806}, \href
  {http://adsabs.harvard.edu/abs/2016MNRAS.462.3146V} {462, 3146}

\bibitem[\protect\citeauthoryear{{V\"olk}}{{V\"olk}}{1989}]{1989lf32}
{V\"olk} H.~J.,  1989, \aap, \href
  {http://adsabs.harvard.edu/abs/1989A%26A...218...67V} {218, 67}

\bibitem[\protect\citeauthoryear{{Williams} et~al.,}{{Williams}
  et~al.}{2016}]{2016williams}
{Williams} W.~L.,  et~al., 2016, \mn@doi [\mnras] {10.1093/mnras/stw1056},
  \href {http://adsabs.harvard.edu/abs/2016MNRAS.tmp..926W} {}

\bibitem[\protect\citeauthoryear{{Wright} et~al.,}{{Wright}
  et~al.}{2010}]{wright10}
{Wright} E.~L.,  et~al., 2010, \mn@doi [\aj] {10.1088/0004-6256/140/6/1868},
  \href {http://adsabs.harvard.edu/abs/2010AJ....140.1868W} {140, 1868}

\bibitem[\protect\citeauthoryear{{Wu}, {Charmandaris}, {Houck},
  {Bernard-Salas}, {Lebouteiller}, {Brandl}  \& {Farrah}}{{Wu}
  et~al.}{2008}]{2008lf31}
{Wu} Y.,  {Charmandaris} V.,  {Houck} J.~R.,  {Bernard-Salas} J.,
  {Lebouteiller} V.,  {Brandl} B.~R.,   {Farrah} D.,  2008, \mn@doi [\apj]
  {10.1086/527288}, \href {http://adsabs.harvard.edu/abs/2008ApJ...676..970W}
  {676, 970}

\bibitem[\protect\citeauthoryear{{York} et~al.,}{{York} et~al.}{2000}]{york00}
{York} D.~G.,  et~al., 2000, \mn@doi [\aj] {10.1086/301513}, \href
  {http://adsabs.harvard.edu/abs/2000AJ....120.1579Y} {120, 1579}

\bibitem[\protect\citeauthoryear{{Yun}, {Reddy}  \& {Condon}}{{Yun}
  et~al.}{2001}]{2001lf9}
{Yun} M.~S.,  {Reddy} N.~A.,   {Condon} J.~J.,  2001, \mn@doi [\apj]
  {10.1086/323145}, \href {http://adsabs.harvard.edu/abs/2001ApJ...554..803Y}
  {554, 803}

\bibitem[\protect\citeauthoryear{{da Cunha}, {Charlot}  \& {Elbaz}}{{da Cunha}
  et~al.}{2008a}]{dacunha08}
{da Cunha} E.,  {Charlot} S.,   {Elbaz} D.,  2008a, \mn@doi [\mnras]
  {10.1111/j.1365-2966.2008.13535.x}, \href
  {http://adsabs.harvard.edu/abs/2008MNRAS.388.1595D} {388, 1595}

\bibitem[\protect\citeauthoryear{{da Cunha}, {Charlot}  \& {Elbaz}}{{da Cunha}
  et~al.}{2008b}]{2008dat5}
{da Cunha} E.,  {Charlot} S.,   {Elbaz} D.,  2008b, \mn@doi [\mnras]
  {10.1111/j.1365-2966.2008.13535.x}, \href
  {http://adsabs.harvard.edu/abs/2008MNRAS.388.1595D} {388, 1595}

\bibitem[\protect\citeauthoryear{{da Cunha}, {Eminian}, {Charlot}  \&
  {Blaizot}}{{da Cunha} et~al.}{2010}]{dacunha10}
{da Cunha} E.,  {Eminian} C.,  {Charlot} S.,   {Blaizot} J.,  2010, \mn@doi
  [\mnras] {10.1111/j.1365-2966.2010.16344.x}, \href
  {http://adsabs.harvard.edu/abs/2010MNRAS.403.1894D} {403, 1894}

\bibitem[\protect\citeauthoryear{{de Jong}, {Klein}, {Wielebinski}  \&
  {Wunderlich}}{{de Jong} et~al.}{1985}]{1985lf1}
{de Jong} T.,  {Klein} U.,  {Wielebinski} R.,   {Wunderlich} E.,  1985, \aap,
  \href {http://cdsads.u-strasbg.fr/abs/1985A%26A...147L...6D} {147, L6}

\bibitem[\protect\citeauthoryear{{van Haarlem} et~al.,}{{van Haarlem}
  et~al.}{2013}]{vanHaarlem2013}
{van Haarlem} M.~P.,  et~al., 2013, \mn@doi [\aap]
  {10.1051/0004-6361/201220873}, \href
  {http://adsabs.harvard.edu/abs/2013A%26A...556A...2V} {556, A2}

\bibitem[\protect\citeauthoryear{{van der Kruit}}{{van der
  Kruit}}{1971}]{1971lf2}
{van der Kruit} P.~C.,  1971, \aap, \href
  {http://cdsads.u-strasbg.fr/abs/1971A%26A....15..110V} {15, 110}

\bibitem[\protect\citeauthoryear{{van der Kruit}}{{van der
  Kruit}}{1973}]{1973lf3}
{van der Kruit} P.~C.,  1973, \aap, \href
  {http://cdsads.u-strasbg.fr/abs/1973A%26A....29..263V} {29, 263}

\makeatother
\end{thebibliography}

\appendix

\section{Evaluation of the flux scale and flux measurements}

\label{app:fluxscale}

\begin{figure}
\begin{center}
\scalebox{0.9}{
\begin{tabular}{c}
\hspace{-4em}\includegraphics[width=11cm,height=11cm,angle=0,keepaspectratio]{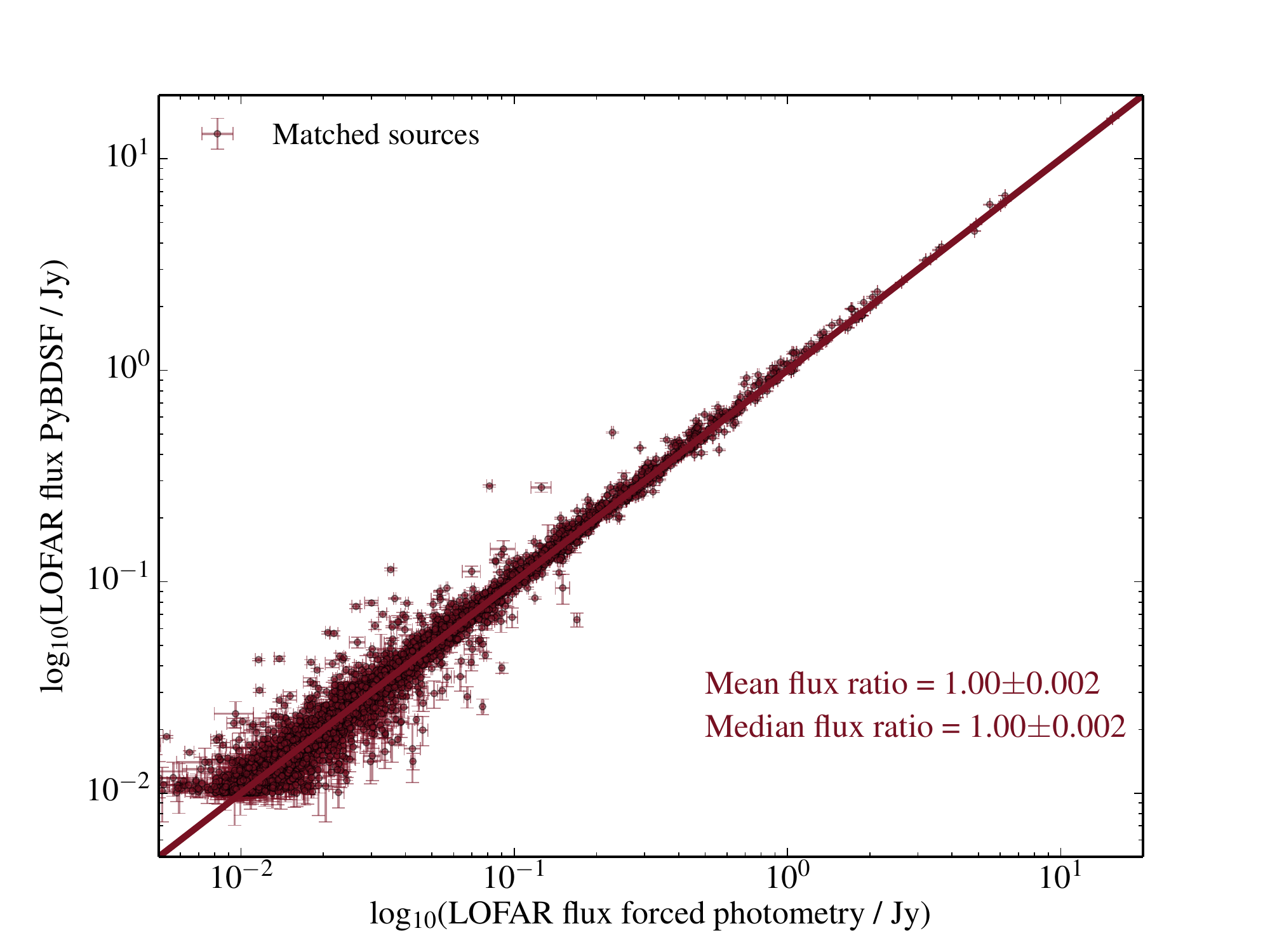}\\
\end{tabular}}
\caption{The distribution of LOFAR-PyBDSF fluxes as a function of LOFAR flux densities of the SFGs measured using the forced photometry method for this work.}
\label{fluxrat1}
\end{center}
\end{figure}

In this Appendix we describe checks that we carried out on the overall flux scale of the LOFAR data and the flux extraction methods used for this specific project. The reprocessing of the data incorporates the `bootstrap' process described by H16 which corrects for LOFAR beam model uncertainties, but it is carried out at an early stage of the calibration using phase-only calibrated images and accordingly we expect slight changes in the flux scale, within the nominal $\sim 10$ per cent errors of the bootstrap procedure. To check that the overall flux scale has not drifted significantly we generated a new PyBDSF \citep{Mohan+Rafferty15} catalogue of the full H-ATLAS field, containing 37,378 sources above $5\sigma$ (more than twice as many as in the H16 analysis). We compared this with the TGSS catalogue described by H16, restricting the comparison to compact sources, with fitted sizes less than 18 arcsec, since the new LOFAR catalogue does not include positional associations for components of extended AGN. We find a median flux ratio (TGSS/LOFAR) of 0.97 and a mean of 1.01 for compact sources with flux density above 20 mJy (chosen to avoid noise limits in the TGSS data which have an rms noise of $\sim 4$ mJy beam$^{-1}$). This is a very similar flux ratio to the overall results of H16 and we conclude that the flux scale has not been negatively affected by the reprocessing. For the processing in this paper we did not make use of the PyBDSF catalogue but used forced photometry at the positions of our target sources, as described in Section \ref{sec:lofar-flux} and \ref{sec:spix}. To verify that this method was giving correct results we also carried out forced photometry at the positions of all isolated, compact point sources (fitted size less than 15 arcsec) with flux densities $S>10$ mJy in the PyBDSF catalogue.A comparison between the flux densities given by PyBDSF and our measured fluxes is shown in Fig. \ref{fluxrat1}, showing good agreement between the two methods. The slightly larger fluxes from PyBDSF can be assumed to be due to the sources at the larger end of the fitted size distribution.

\begin{figure}
\begin{center}
\scalebox{0.9}{
\begin{tabular}{c}
\hspace{-4em}\includegraphics[width=11cm,height=11cm,angle=0,keepaspectratio]{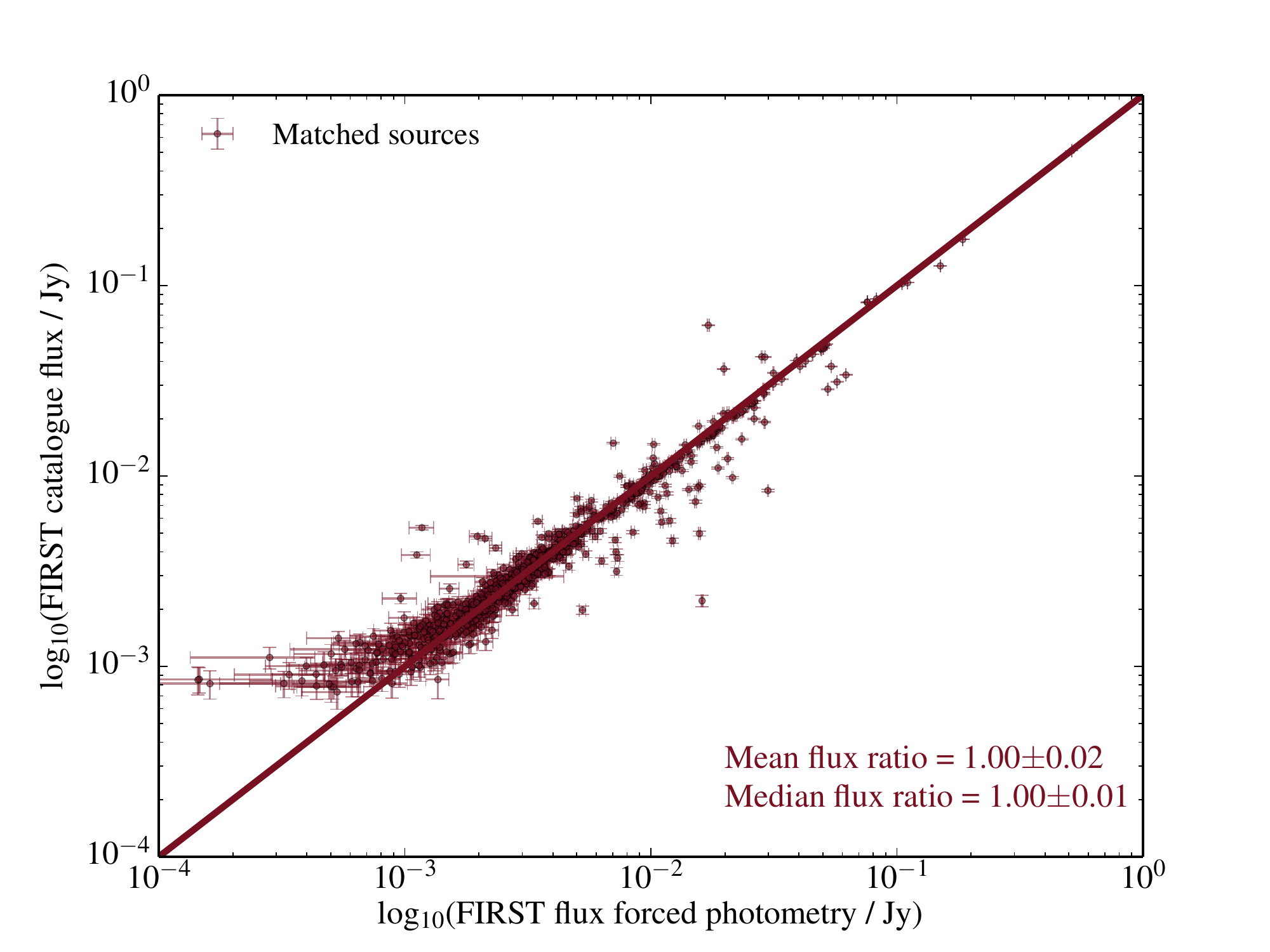}\\
\end{tabular}}
\caption{The distribution of FIRST fluxes as a function of LOFAR fluxes measured using the forced photometry method for this work.}
\label{fluxrat2}
\end{center}
\end{figure}

The same evaluation was carried out for our FIRST measurements: we cross-matched the source catalogue with the FIRST survey data in the field. We then filtered out sources with FIRST sidelobe probability (i.e. probability of being an artefact) $>0.05$. We restricted the crossmatching to compact LOFAR sources (fitted size less than 10 arcsec) with well-determined positions (nominal positional error less than 5 arcsec). Fig. \ref{fluxrat2} shows these matches and the agreement between flux densities is clear. We conclude that we measure flux densities reliably using the aperture photometry method.

\section{Comparison of star formation rate indicators}

\label{app:magphys}

\begin{figure}
\begin{center}
\scalebox{0.9}{
\begin{tabular}{c}
\hspace{-4em}\includegraphics[width=11cm,height=11cm,angle=0,keepaspectratio]{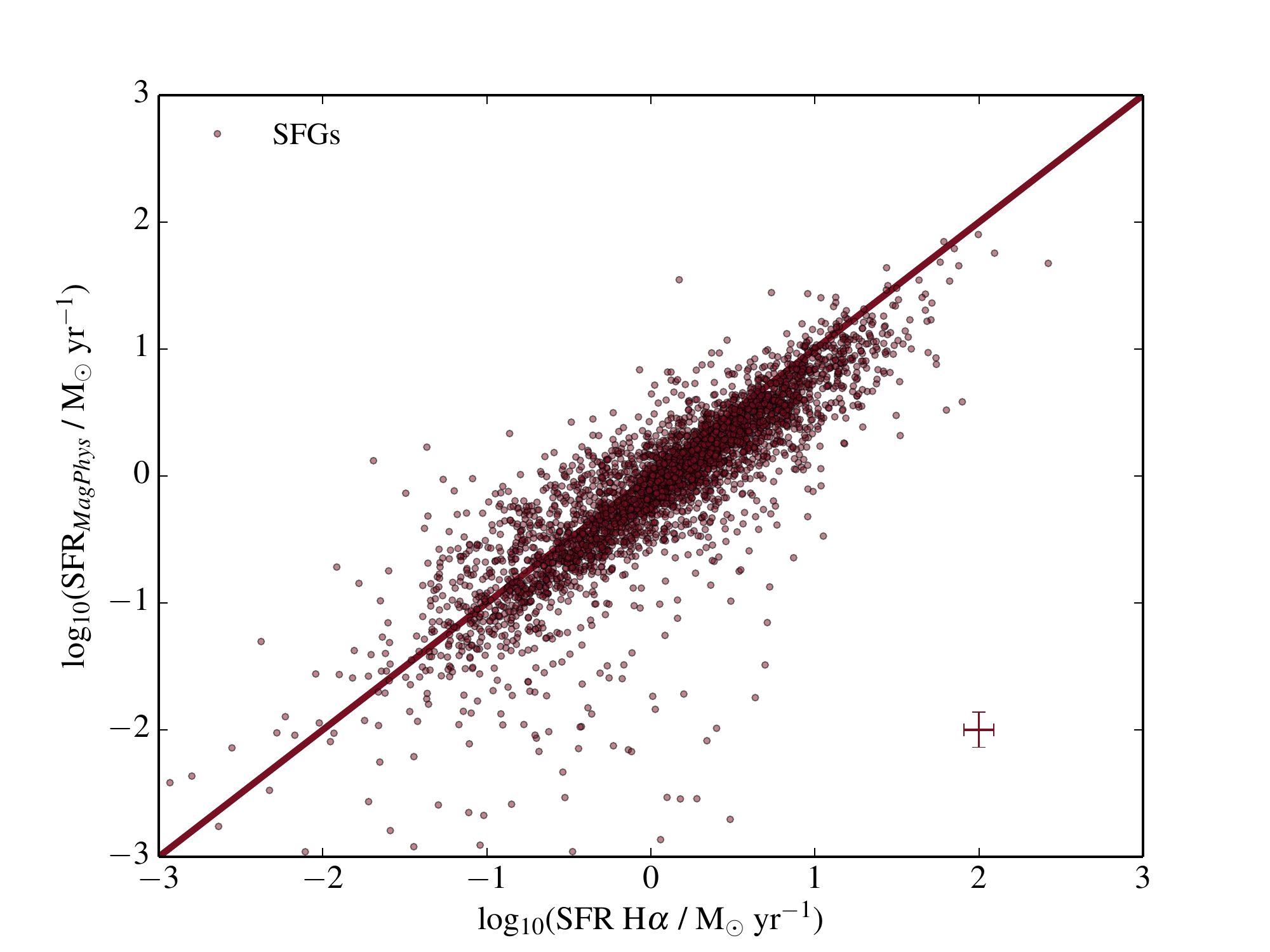}\\
\hspace{-4em}\includegraphics[width=11cm,height=11cm,angle=0,keepaspectratio]{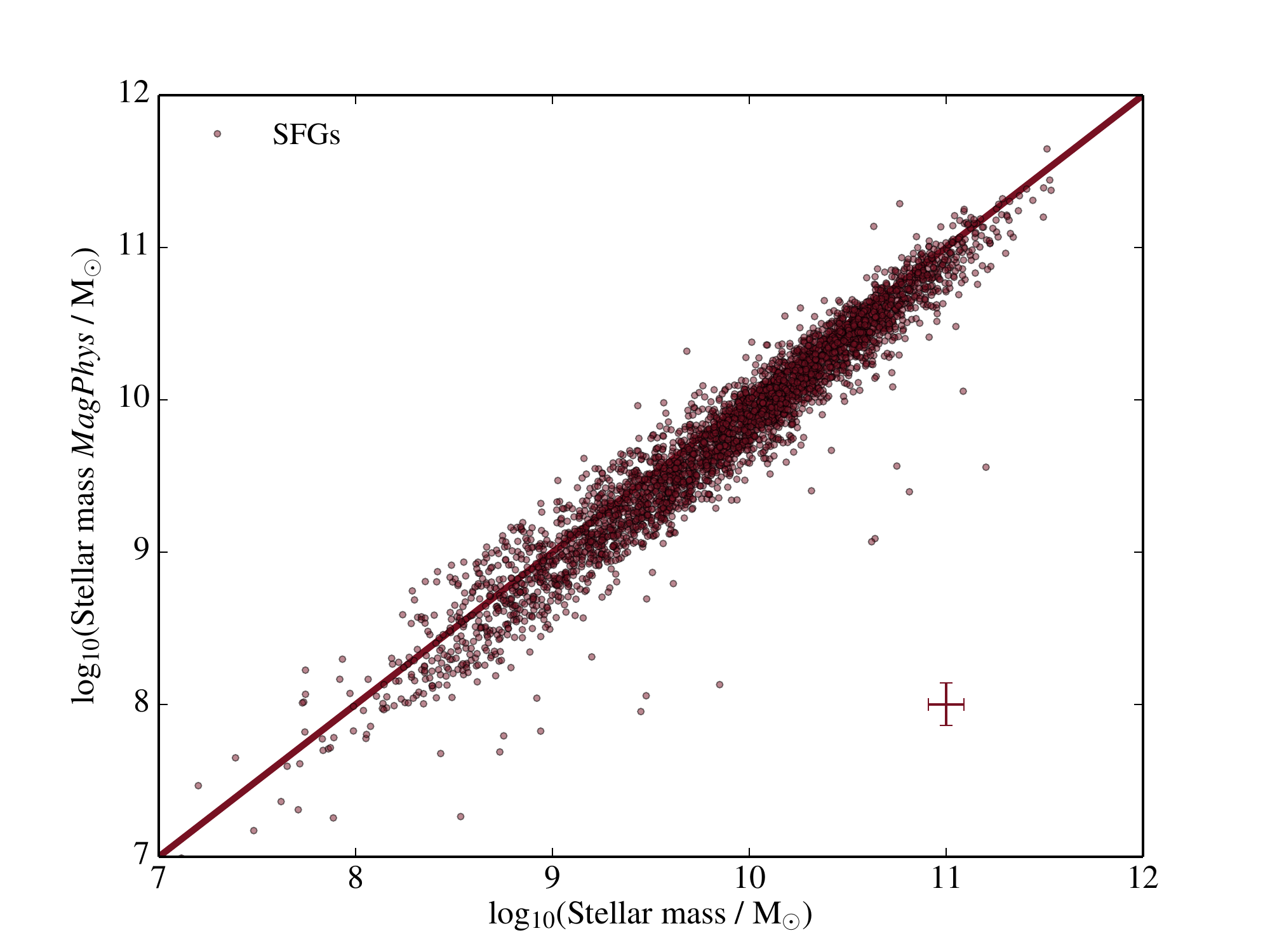}\\
\end{tabular}}
\caption{Top: The distribution of \magphys\ best-fit estimates of SFRs of SFGs in the sample versus their H${\alpha}$-SFRs. The solid black line shows the one to one relation in both panels. Bottom: The distribution of \magphys\ best-fit estimates of stellar masses of SFGs in the sample versus their stellar masses given in the MPA-JHU catalogue.}
\label{sfr-sfr}
\end{center}
\end{figure}

 SFR measurements for galaxies, calculated using the H${\alpha}$ emission line, corrected for dust attenuation and fibre aperture effects, are also available in the MPA-JHU data base \citep{2004ref8}. As a sanity check on our results we compare SFRs of SFGs in the sample derived by the two different methods. In the top panel of Fig. \ref {sfr-sfr} we show \magphys\ the best-fit estimates of SFRs from \magphys\ and H$\alpha$-SFRs. In the bottom panel of the same figure we compare best-fit estimates of stellar masses of SFGs from \magphys\ and stellar masses (calculated from fits to the photometry) provided in the MPA-JHU catalogue. We see that in general H${\alpha}$-SFR results are in good agreement with the \magphys\ best-fit estimates, with perhaps a slight tendency for the \magphys\ values to be lower.

There are various differences in the derivations of these SFRs which might cause the residual differences between the values of SFRs of galaxies in the sample estimated using the two methods. First of all, the correction of dust attenuation was calculated differently in the two different SFR estimation methods. For the H$\alpha$-SFRs dust attenuation was estimated from H$\alpha$/H$\beta$ assuming a fixed unattenuated Case B ratio \citep{2004ref8}. As stated by \citet{2004ref8}, using a single conversion factor is not fully correct but is a reasonable approximation. On the other hand, \magphys\ uses a more complicated procedure to estimate dust attenuation \citep[see][for details]{2008dat5}. Also, H$\alpha$ SFRs suffer from aperture effects. Furthermore, the extragalactic dust law is not fully known. Although for both H${\alpha}$-SFRs and \magphys$-$SFRs the same dust law \citep{charlot00} was used, the extragalactic dust law might vary as a function of galaxy properties. Recently, Wang et al. (2016) compared different SFRs estimated by different methods and found around $\sim$ 0.2 dex difference between H${\alpha}$-SFRs and those derived from the far-IR and UV data, and they also observed variations as a function of galaxy parameters. Similarly, \citet{2016davies} compared various SFR estimators including H$\alpha$ and found differences between different methods. Taking into account all the complexity involved in the derivation of SFRs, a small systematic offset between H${\alpha}$-SFR and \magphys$-$SFR is not unexpected. There is no known right answer, but we adopt the \magphys\ results for all of our analyses here, since this method applies a physically meaningful obscuration correction via SED modelling.

\end{document}